\documentclass[traditabstract]{aa}

\usepackage{graphicx}
\usepackage{amssymb}
\usepackage{amsmath}
\usepackage{url}
\usepackage{natbib}
\bibpunct{(}{)}{;}{a}{}{,}

\newcommand{\onefig}[3]{%
  \begin{figure}%
    \centerline{\resizebox{\hsize}{!}{\includegraphics*{#1}}}%
    \caption{#3}\label{#2}%
  \end{figure}%
}
\newcommand{\onelargefig}[3]{%
  \begin{figure*}%
    \centerline{\resizebox{\hsize}{!}{\includegraphics*{#1}}}%
    \caption{#3}\label{#2}%
  \end{figure*}%
}
\newcommand{\twofig}[4]{%
  \begin{figure*}%
    \centerline{\resizebox{\hsize}{!}{\includegraphics*{#1} \,%
        \includegraphics*{#2}}}%
    \caption{#4}\label{#3}%
  \end{figure*}%
}
\newcommand{\threefig}[5]{%
  \begin{figure*}%
    \centerline{\resizebox{\hsize}{!}{\includegraphics*{#1} \,%
        \includegraphics*{#2} \,%
        \includegraphics*{#3}}}%
    \caption{#5}\label{#4}%
  \end{figure*}%
}

\newcommand{\sect}[1]{Sect.~\ref{#1}}

\newcommand{\fig}[1]{Fig.~\ref{#1}}

\newcommand{\eq}[1]{Eq.~(\ref{#1})}

\newcommand\lmax{\ell_{\mathrm{max}}}

\newcommand\lm{\ell m}
\newcommand\alm{$a_{\ell m}$}
\newcommand\order[1]{${{\cal O}\! \left( #1 \right)}$}
\newcommand\beq{\begin{equation}}
\newcommand\eeq{\end{equation}}

\begin{document}

\title{Likelihood, Fisher information, and systematics of cosmic
  microwave background experiments}

\titlerunning{Likelihood, Fisher information, and systematics of CMB
  experiments}

\author{Franz Elsner\inst{1}
  \and
  Benjamin D. Wandelt\inst{1,2}}

\institute{Institut d'Astrophysique de Paris, UMR 7095, CNRS -
  Universit\'e Pierre et Marie Curie (Univ Paris 06), 98 bis blvd
  Arago, 75014 Paris, France\\
  \email{elsner@iap.fr}
  \and
  Departments of Physics and Astronomy, University of Illinois at
  Urbana-Champaign, Urbana, IL 61801, USA}

\date{Received \dots / Accepted \dots}

\abstract{Every experiment is affected by systematic effects that
  hamper the data analysis and have the potential to ultimately
  degrade its performance. In the case of probes of the cosmic
  microwave background (CMB) radiation, a minimal set of issues to
  consider includes asymmetric beam functions, correlated noise, and
  incomplete sky coverage. Presuming a simplified scanning strategy
  that allows for an exact analytical treatment of the problem, we
  study the impact of systematic effects on the likelihood function of
  the CMB power spectrum. We use the Fisher matrix, a measure of the
  information content of a data set, for a quantitative comparison of
  different experimental configurations. In addition, for various
  power spectrum coefficients, we explore the functional form of the
  likelihood directly, and obtain the following results: The
  likelihood function can deviate systematically from a Gaussian
  distribution up to the highest multipole values considered in our
  analysis. Treated exactly, realistic levels of asymmetric beam
  functions and correlated noise do not by themselves decrease the
  information yield of CMB experiments nor do they induce noticeable
  coupling between multipoles. Masking large fractions of the sky, on
  the other hand, results in a considerably more complex correlation
  structure of the likelihood function. Combining adjacent power
  spectrum coefficients into bins can partially mitigate these
  problems.}


\maketitle

\section{Introduction}
\label{sec:intro}

Owing to its sensitive shape dependence on the most fundamental
cosmological parameters, the power spectrum of the cosmic microwave
background (CMB) radiation is one of the most important probes of the
properties of the early universe \citep{2004IJTP...43..623M}. As a
consequence, a large number of experiments have been designed to
improve the observational constraints on the CMB anisotropies
\citep[e.g.,][]{1992ApJ...396L...1S, 2002ApJ...571..604N,
  2003ApJS..148....1B, 2004ApJ...600...32K, 2011ApJ...739...52D,
  2011ApJ...743...28K, 2011A&A...536A...1P}. With the increasing
precision of observations, the demands on the numerical tools used for
data analysis became more and more stringent. Given the data obtained
with the Planck satellite, for example, the temperature power spectrum
will be cosmic variance limited from the largest scales up to a
multipole moment of about $\ell \approx 2500$
\citep{2005planck_bluebook}. To take full advantage of this wealth of
information, a precise understanding and thorough modeling of the data
properties are necessary.

The Fisher information matrix is ubiquitous in statistics. Used in the
context of the Cramer-Rao bound, it permits a quantitative assessment
of an experiment's ability to constrain a set of
parameters. Furthermore, it also allows drawing conclusions about
parameter degeneracies. Owing to its predictive power, the Fisher
matrix has been widely used to characterize and optimize experiments
in various fields of astronomy \citep[for example, in cosmology,
  e.g.,][]{1999ApJ...514L..65H, 2005PhRvD..71h4025B,
  2005MNRAS.360L..82R, 2008ApJ...686L...1L}. Since the computational
costs associated with its numerical calculation are prohibitively
large, such studies have not been realized for general high-resolution
CMB experiments. \citet{2012A&A...540L...6E} introduced a method for
an efficient yet exact calculation of the Fisher matrix and of the
full likelihood in only \order{\lmax^4} operations. This mathematical
framework enables a quantitative study of several systematic effects
on CMB power spectrum constraints. In this work, we focus on common
systematics like asymmetric beam functions, correlated noise, partial
sky coverage, and the effect of binning. For different experimental
setups, we analyze the correlation structure of the power spectrum
coefficients and explore the shape of the likelihood function itself.

Cosmological parameter estimation from a power spectrum is commonly
performed by means of Monte Carlo sampling techniques
\citep[e.g.][]{2000astro.ph..6401C, 2002PhRvD..66j3511L}. Owing to the
costs associated with the numerical evaluation of the full likelihood
function given the data map, exact methods are restricted to the
lowest multipole moments, $\lmax \la 50$
\citep[e.g.][]{2007ApJS..170..288H, 2009MNRAS.400..463G}. Beyond that,
the likelihood function is usually written in terms of the power
spectrum coefficients themselves, a considerably less demanding
problem. To do so, the function has to be approximated by, e.g., a
multivariate Gaussian distribution \citep[as done for the analysis of
  the Atacama Cosmology Telescope data,][]{2011ApJ...739...52D}, or
some other choice, such as a combination of a Gaussian and an offset
log-normal distribution \citep[as done by the WMAP
  team,][]{2003ApJS..148..195V}. Using our methods, we are able to
give examples of generic features of the true underlying likelihood
function which should be reproduced by these approximations to avoid
introducing systematic biases in the cosmological parameter
determinations.

Since our method allows an exact modeling of the data in its full
complexity, we can address a new question about how different
instrumental and experimental properties affect the information
content of a CMB data set: rather than asking how systematics affect
results if they are not taken into account optimally (see, e.g., the
discussion of correlated noise in \citealt{1998A&AS..127..555D,
  1999A&AS..140..383M}, asymmetric beams in
\citealt{2003A&A...409..423B, 2011ApJS..193....5M}, and partial sky
coverage in \citealt{2001ApJ...561L..11S, 2001PhRvD..64h3003W,
  2002ApJ...567....2H}), we can determine to what extent these
systematics fundamentally degrade power spectrum constraints even after
an analysis that takes their effects into account optimally. In all
the analyses we present in this paper, the instrumental
characteristics of the simulated experiment (such as the noise model
and the asymmetric beam maps) are assumed to be known exactly. To make
this assumption true for realistic experiments requires a careful
calibration strategy. Therefore, the results we discuss should be seen
as hard lower limits on the effects of systematics on CMB power
spectrum inferences.

The paper is organized as follows. In \sect{sec:torus}, we briefly
review the mathematical background used for our study. We then assess
the impact of important experimental systematics on the likelihood
function (\sect{sec:fisher}). Finally, we summarize our findings in
\sect{sec:summary}.

\section{The likelihood analysis on the ring torus}
\label{sec:torus}

Consider an arbitrary CMB experiment with Gaussian noise, the
likelihood function ${\cal L}$ of the data $d$ is
\begin{multline}
\label{eq:likelihood}
{\cal L}(C_{\ell} \vert d) = \frac{1} {\sqrt{ \vert 2 \pi
    (\mathbf{S}(C_\ell) + \mathbf{N}) \vert}} \\
\times \exp \left[ {-\frac{1}{2} \, d^{\dagger} (\mathbf{S}(C_\ell) +
    \mathbf{N})^{-1} d} \right] \, ,
\end{multline}
where we have introduced the signal covariance matrix, $\mathbf{S}$,
as a function of the CMB power spectrum coefficients $C_{\ell}$. The
noise properties are characterized by the noise covariance matrix,
$\mathbf{N}$.

In the most general case, the evaluation of \eq{eq:likelihood} takes
\order{N^3} operations, where $N$ is the number of pixels in the
survey \citep{1999PhRvD..59b7302B}. To enable a numerically feasible
analysis, we only consider experiments with an idealized scanning
strategy, where the sky is observed on iso-latitude circles. The
mathematical background of the algorithm is explained in detail in
\citet{2003PhRvD..67b3001W} and \citet{2012A&A...540L...6E}. Following
their approach, we then map the time-ordered data (TOD) onto the ring
torus. We visualize this duality in
\fig{fig:scanning_strategy}. Casting the equations into Fourier space
allows us to take advantage of the periodicity of the manifold -- the
expression for the likelihood function simplifies.

\onefig{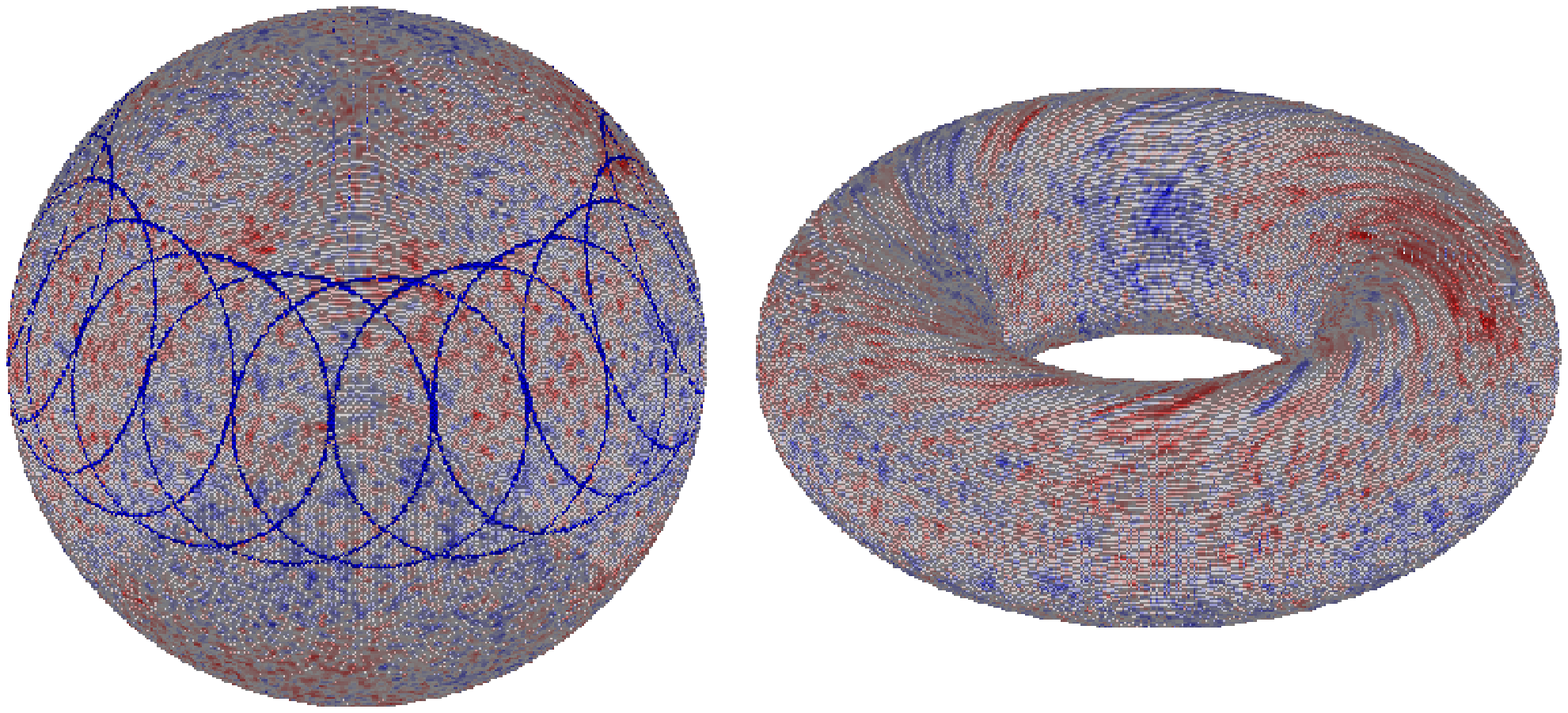}{fig:scanning_strategy}
{Idealizing the scanning strategy leads to additional symmetries.
  Observing the sky on iso-latitude rings with a regular spacing in
  azimuthal direction (\emph{left panel}) allows to map the resulting
  data structure onto a ring torus (\emph{right panel}). Then, the
  formulation of the likelihood function in Fourier basis simplifies.}

To obtain the signal covariance matrix in Fourier basis, we specify
the noise-free sky temperature as a function of the spherical harmonic
coefficients of a signal map, \alm,
\beq
\label{eq:signal_map}
T^{S}_{r p} = \sum_{\ell} a_{\ell r} \, d^{\ell}_{r
  p}(\theta_{\mathrm{s}}) \, X_{\ell p} \, .
\eeq
Here, the index $p$ specifies the Fourier modes in the in-ring
direction, and $r$ labels the index for the cross-ring direction. In
\eq{eq:signal_map}, we make use of the definition of the Wigner
rotation matrix to introduce the real quantity $d$,
\beq
D^{\ell}_{m m'}(\phi_2,\theta,\phi_1) = \mathrm{e}^{-\mathrm{i} m
  \phi_2} \, d^{\ell}_{m m'}(\theta) \, \mathrm{e}^{-\mathrm{i} m'
  \phi_1} \, ,
\eeq
which we evaluate at the latitude of the experiment's spin axis,
$\theta = \theta_{\mathrm{s}}$. We also apply the rotated beam $X$
according to
\beq
X_{\ell m} = \sqrt{\frac{2 \ell + 1}{4 \pi}} \sum_{m'} d^\ell_{m
  m'}(\theta_{\mathrm{o}}) \, b_{\ell m'}^{\ast} \, ,
\eeq
where the Wigner small $d$ matrix is now computed at the opening angle
of the scanning circles, $\theta_{\mathrm{o}}$. The spherical harmonic
expansion coefficients of the beam pattern, $b_{\lm}$, are taken into
account exactly and need not to be azimuthally averaged. The signal
covariance matrix $\langle T^{S}_{r p} T^{S \, \ast}_{r' p'} \rangle$
then simplifies,
\beq
\mathbf{S}_{r p r' p'} = \delta_{r r'} N^2 \sum_{\ell} C_{\ell} \,
d^{\ell}_{r p} X_{\ell p} \, d^{\ell}_{r' p'} X_{\ell p'}^{\ast} \, .
\eeq
In Fourier space, $\mathbf{S}$ shows a block diagonal structure.

We now specify the noise correlations. For an experiment with
stationary noise, we can fully characterize its properties by a power
spectrum $P(k)$ in the TOD domain. This ansatz is general enough to
include the exact treatment of, e.g., $1/f^2$-type noise, a common
systematic in real-world experiments. The noise covariance matrix
$\langle T^{N}_{r p} T^{N \, \ast}_{r' p'} \rangle$ now reduces to
\begin{multline}
\mathbf{N}_{r p r' p'} = \delta_{r r'} \frac{1}{N_{\mathrm{r}}
  N_{\mathrm{p}}^2} \sum_{m, m' = 0}^{N_{\mathrm{p}} - 1}
\mathrm{e}^{-\frac{2 \pi \mathrm{i}}{N_{\mathrm{p}}}
  (p m - p' m')} \\
\times \sum_{\Delta = 0}^{N_{\mathrm{r}} - 1} \mathrm{e}^{-\frac{2
    \pi \mathrm{i}}{N_{\mathrm{r}}} \Delta r} C(\Delta, m, m') \, ,
\end{multline}
where $N_{\mathrm{p}}$ is the number of pixels per ring, and
$N_{\mathrm{r}}$ is the number of rings in the data set. We introduced
an auxiliary function $C(\Delta, m, m')$ according to
\beq
C(\Delta, m, m') = \sum_{k=0}^{N-1} P(k) \, e^{-\frac{2\pi \mathrm{i}}
  {N} k (N_{\mathrm{r}} \Delta + m - m')} \, .
\eeq
In a Fourier space representation, we also find the noise covariance
matrix $\mathbf{N}$ to be block diagonal.

We now review an efficient scheme to calculate the Fisher information
matrix. Defined as the covariance of the score function, from
\eq{eq:likelihood}, we obtain \citep{1997PhRvD..55.5895T}
\begin{align}
 \label{eq:fisher}
 F_{\ell_{1} \ell_{2}} &= - \left \langle \frac{\partial^2 \ln {\cal
     L}}{\partial C_{\ell_{1}} \partial C_{\ell_{2}}} \right \rangle
 \nonumber \\
 &= \frac{1}{2} \, \mathrm{tr} \left[ \mathbf{C}^{-1} \mathbf{P}^{\ell_{1}}
   \mathbf{C}^{-1} \mathbf{P}^{\ell_{2}} \right] \, ,
\end{align}
where $\mathbf{C} = \mathbf{S} + \mathbf{N}$, and $\mathbf{P}^{\ell} =
\partial \mathbf{C} / \partial C_{\ell}$. For the derivative of the
covariance matrix with respect to a power spectrum coefficient, we
find for each matrix block $r$,
\begin{align}
\label{eq:cov_deriv}
\left( \frac{\partial \mathbf{C}_{r}}{\partial C_{\ell}} \right)_{p \,
  p'}&= \left( \frac{\partial \mathbf{S}_{r}}{\partial C_{\ell}}
\right)_{p \, p'} \nonumber \\
&= N^2 \, d^{\ell}_{r p} X_{\ell p} \, d^{\ell}_{r p'} X_{\ell p'}^{\ast}
\nonumber \\
&= q^{\ell}_{r p} \, q^{\ell \, \ast}_{r p'} \, .
\end{align}
On the ring torus, each block in \eq{eq:cov_deriv} is a rank one
object and can be decomposed according to $\mathbf{P}^{\ell}_{r} =
q^{\ell}_{r} \, q^{\ell \, \dagger}_{r}$. The calculation of the
Fisher matrix therefore simplifies,
\beq
\label{eq:fisher_torus}
F_{\ell_{1} \ell_{2}} = \frac{1}{2} \sum_{r} \vert q^{\ell_{1} \,
  \dagger}_{r} \mathbf{C}^{-1}_{r} q^{\ell_{2}}_{r} \vert^2 \, .
\eeq

For completeness, we also provide the expression for the score
function on the ring torus,
\beq
\label{eq:score_torus}
\frac{\partial \ln {\cal L}}{\partial C_{\ell}} = \frac{1}{2} \sum_{r}
\left( \vert d^{\dagger} \mathbf{C}^{-1}_{r} q^{\ell}_{r} \vert^2 -
q^{\ell \, \dagger}_{r} \mathbf{C}^{-1}_{r} q^{\ell}_{r} \right) \, ,
\eeq
which, just as the Fisher matrix, can also be calculated efficiently
owing to its block diagonal structure.

To evaluate the equations numerically, we developed a hybrid
OpenMP/MPI implementation of the algorithm. The intrinsic parallelism
of the method is reflected in a good scaling behavior for up to
$10\,000$ CPU cores and more.

\section{The role of systematic effects}
\label{sec:fisher}

For a detailed study of various effects, we directly compare a
reference simulation, with a simple setup, to a second data
realization where we included another layer of complexity. Although we
only consider an idealized experiment with simplified scanning
strategy, this concurrent analysis allows us to draw conclusions about
the general relevance of the effect under study.

For our benchmark simulation, we used the WMAP7+BAO+H0 cosmological
parameters \citep{2011ApJS..192...18K} to generate a CMB temperature
map, up to a maximum multipole moment of $\lmax = 1023$. According to
\eq{eq:signal_map}, we sampled the signal regularly on $2048^2$ points
on the sphere, i.e., $N_{\mathrm{p}} = N_{\mathrm{r}} = 2 \lmax + 1 =
2048$. The scanning rings had an opening angle of $\theta_{\mathrm{o}}
= 40 \degr$ at a latitude of $\theta_{\mathrm{s}} = 70 \degr$, leading
to a sky coverage of about $f_{\mathrm{sky}} = 60~\%$. For the beam
function, we used a symmetric Gaussian profile with a full width at
half maximum (FWHM) of $12 \arcmin$. In the noise model considered
here, we parameterized the noise power spectrum in the TOD domain by
means of a simple function,
\beq
\label{eq:noise}
P(k) = \sigma^2 \left[ 1 + \left( k_{\mathrm{knee}} / k
  \right)^{\alpha}\right] \, ,
\eeq
allowing to include $1/f^{\alpha}$-type correlations. For the
benchmark, we added a noise realization drawn from a white noise power
spectrum (i.e., $\alpha = 0$) with a fixed amplitude $\sigma =
10$~mK. At this level, the experiment is signal dominated up to the
highest multipole moment considered.

We base our analysis on a twofold approach: on the evaluation of the
likelihood function, and on the Fisher information matrix. Starting
from the fiducial CMB power spectrum, we use \eq{eq:fisher_torus} to
calculate the Fisher matrix for a given experimental setup. The
marginalized 1-$\sigma$ error bars of the power spectrum coefficients
are then given by $\sigma_{C_{\ell}} = \sqrt{(F^{-1})_{\ell \, \ell}}$,
though, strictly speaking, this relation only holds for multivariate
Gaussian distributions. We define the normalized correlation matrix of
the power spectrum coefficients according to
\beq
\label{eq:correlation}
K_{\ell_{1} \ell_{2}} = (F^{-1})_{\ell_{1} \ell_{2}} /
\sqrt{(F^{-1})_{\ell_{1} \ell_{1}} (F^{-1})_{\ell_{2} \ell_{2}}} -
\delta_{\ell_{1} \ell_{2}} \, ,
\eeq
where we forced the diagonal elements to zero. As an independent test,
we probe the shape of the likelihood function directly. To this end,
we start from the fiducial power spectrum and find the likelihood peak
by means of Newton-Raphson iterations \citep{1999ApJ...510..551O},
\beq
\label{eq:iteration}
\hat{C}^{i+1}_{\ell} = \hat{C}^{i}_{\ell} - \frac{1}{2} \sum_{\ell'}
\left( F^{-1} \right)_{\ell \, \ell'} \frac{\partial \ln {\cal
    L}}{\partial C_{\ell'}} \, .
\eeq
We then calculate likelihood slices by varying one power spectrum
coefficient around the best fit value, while keeping all others
fixed. This approach explores the functional form of the likelihood
without the need to rely on approximations.

\subsection{The benchmark}

In \fig{fig:bench_fisher}, we show the estimated maximum likelihood
power spectrum of the reference simulation, $\hat{C}_{\ell}$, and
compare it to the input we adopted to synthesize the CMB
anisotropies. Approximating the likelihood as Gaussian around the
peak, we can calculate the goodness of fit using the Fisher matrix $F$
according to
\beq
\chi^2 = (\hat{C}_{\ell} - C^{\mathrm{\, Input}}_{\ell})^{\dagger} F
\, (\hat{C}_{\ell} - C^{\mathrm{\, Input}}_{\ell}) \, .
\eeq
Since monopole and dipole are excluded from the fit, we set
$N_{\mathrm{dof}} = \lmax - 1$, and obtain $\chi^2/N_{\mathrm{dof}} =
1.09$. We conclude that the fiducial model is consistently
recovered. The Fisher matrix, depicted in the right hand panel of
\fig{fig:bench_fisher}, exhibits a non-trivial shape with noticeable
deviations from a simple diagonal matrix.

To characterize the correlation structure of the power spectrum, we
plot the covariance matrix in \fig{fig:bench_correlation}. In this
diagram, rows or columns at low multipole moments appear to be
dominating. However, this effect is the result of the approximate
$\ell^{-2}$ scaling of the CMB power spectrum. We therefore use
\eq{eq:correlation} to normalize the covariance matrix for an unbiased
assessment of the correlation structure and find no evidence for an
enhanced coupling of small and large scale perturbations\footnote{We
  note that pseudo-$C_{\ell}$ power spectrum estimators are known to
  introduce spurious correlations between small and large multipole
  moments \citep[see, e.g.,][]{2004MNRAS.349..603E}.}. Instead, the
normalized correlation matrix is dominated by its diagonal elements
which we illustrate at low, intermediate, and high multipoles in the
right hand panel of \fig{fig:bench_correlation}. Obviously, a power
spectrum coefficient $C_{\ell}$ shows a substantial negative
correlation at the 10~\% level with its direct neighbors $C_{\ell \pm
  1}$ and $C_{\ell \pm 2}$. Longer-ranging correlations are suppressed
to below 1~\%. As the signal-to-noise ratio (S/N) is large in the
entire multipole range considered here, the evolution of the
correlation coefficients towards higher $\ell$ is negligible.

It has been noted before, that the likelihood function is strongly
non-Gaussian at the lowest multipole moments
\citep[e.g.,][]{1998PhRvD..57.2117B, 2004PhRvD..70h3511W}. As a
consequence, exact sampling methods are commonly used for the data
analysis in this regime \citep{2009ApJS..180..225H,
  2011ApJS..192...14J}. However, and in contrast to previous claims in
literature, we also find a noticeable deviation from Gaussianity at
the highest multipole values. We illustrate the shape of the
likelihood function for large, intermediate, and small scale
perturbations in \fig{fig:bench_like}. For all power spectrum
coefficients, a fit with a simple Gaussian distribution leaves
residuals at the level of several percent. Combining the individual
errors for each multipole moment, the total error on the likelihood
function can be substantial if it is modeled by a simple multivariate
Gaussian.

\twofig{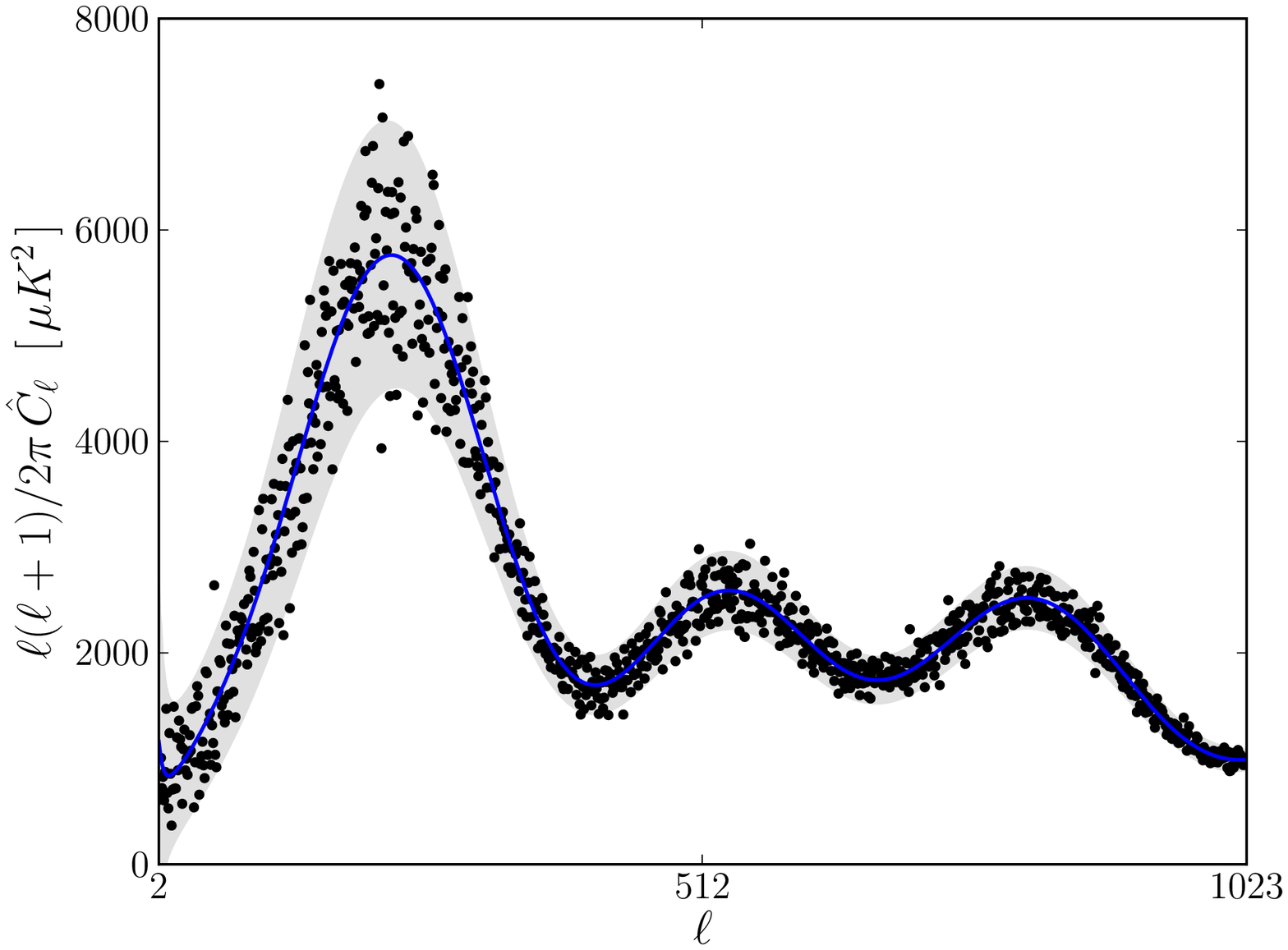}{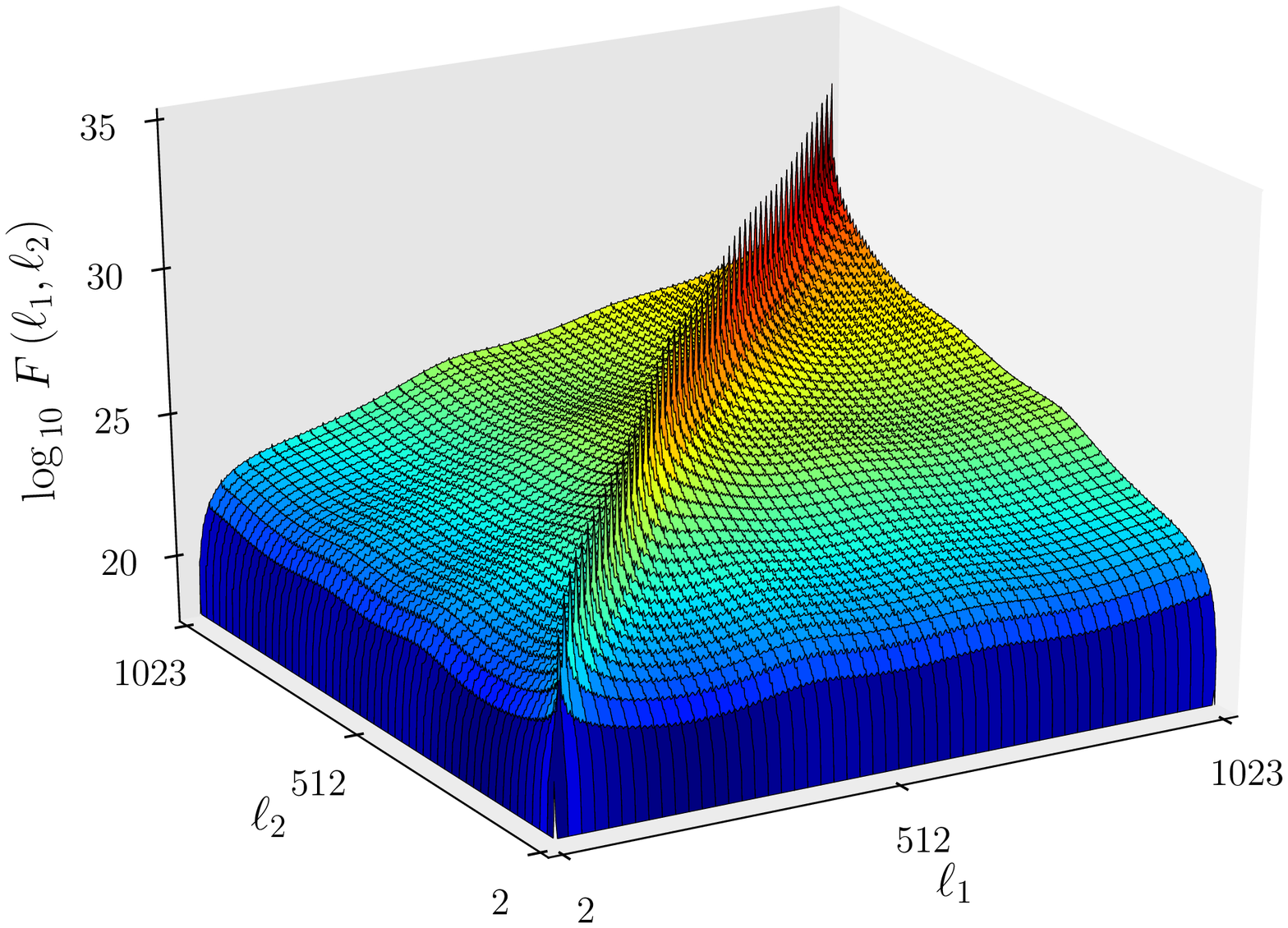}{fig:bench_fisher}
{\emph{Left panel:} The result of an exact maximum likelihood power
  spectrum estimation (\emph{filled circles}) shows that the input
  power spectrum can be consistently recovered (\emph{solid
    line}). The gray area indicates the 2-$\sigma$ confidence region
  as derived from the Fisher matrix. \emph{Right panel:} The Fisher
  matrix of the CMB power spectrum coefficients cannot be approximated
  well by a diagonal matrix.}

\twofig{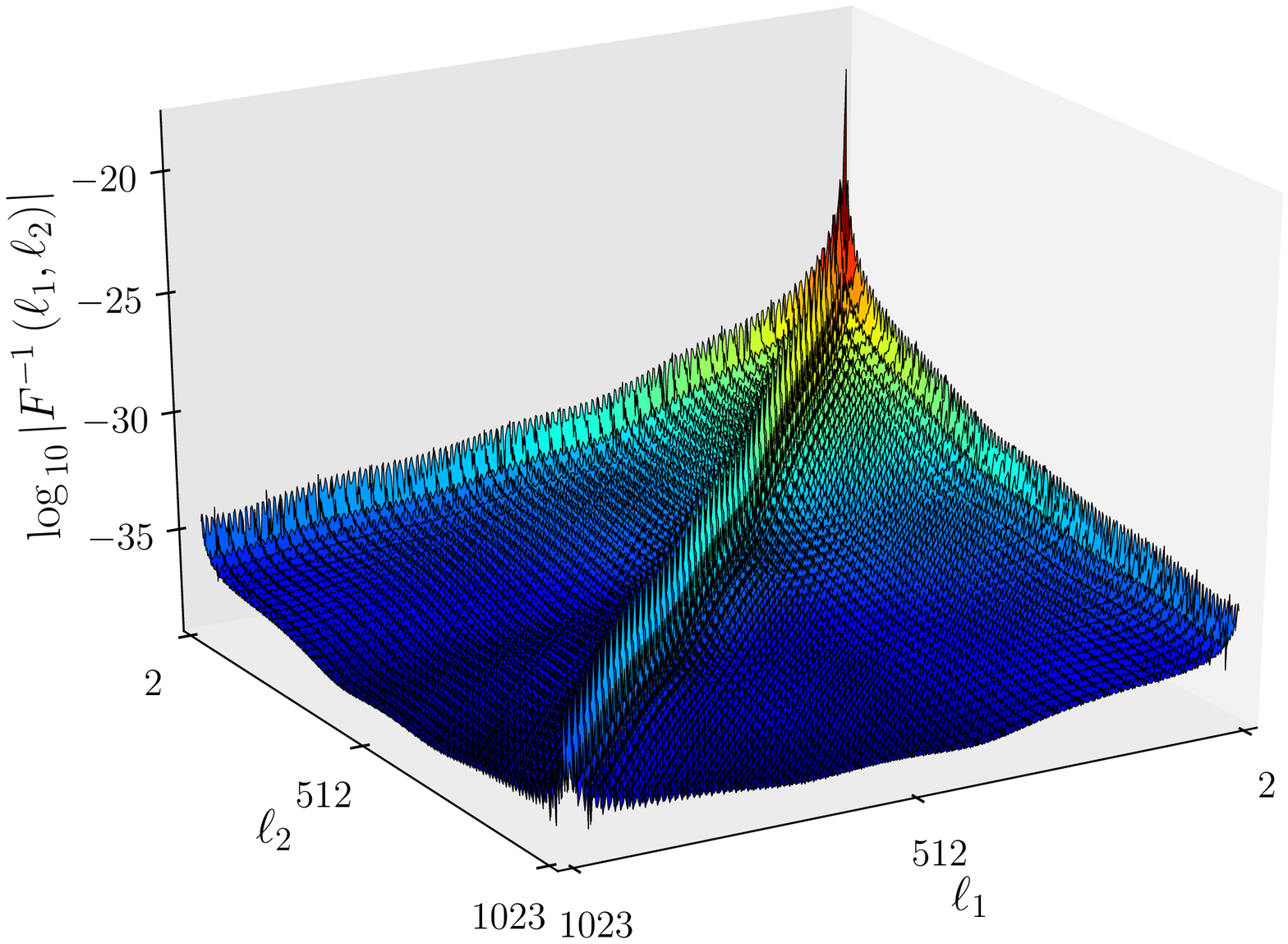}{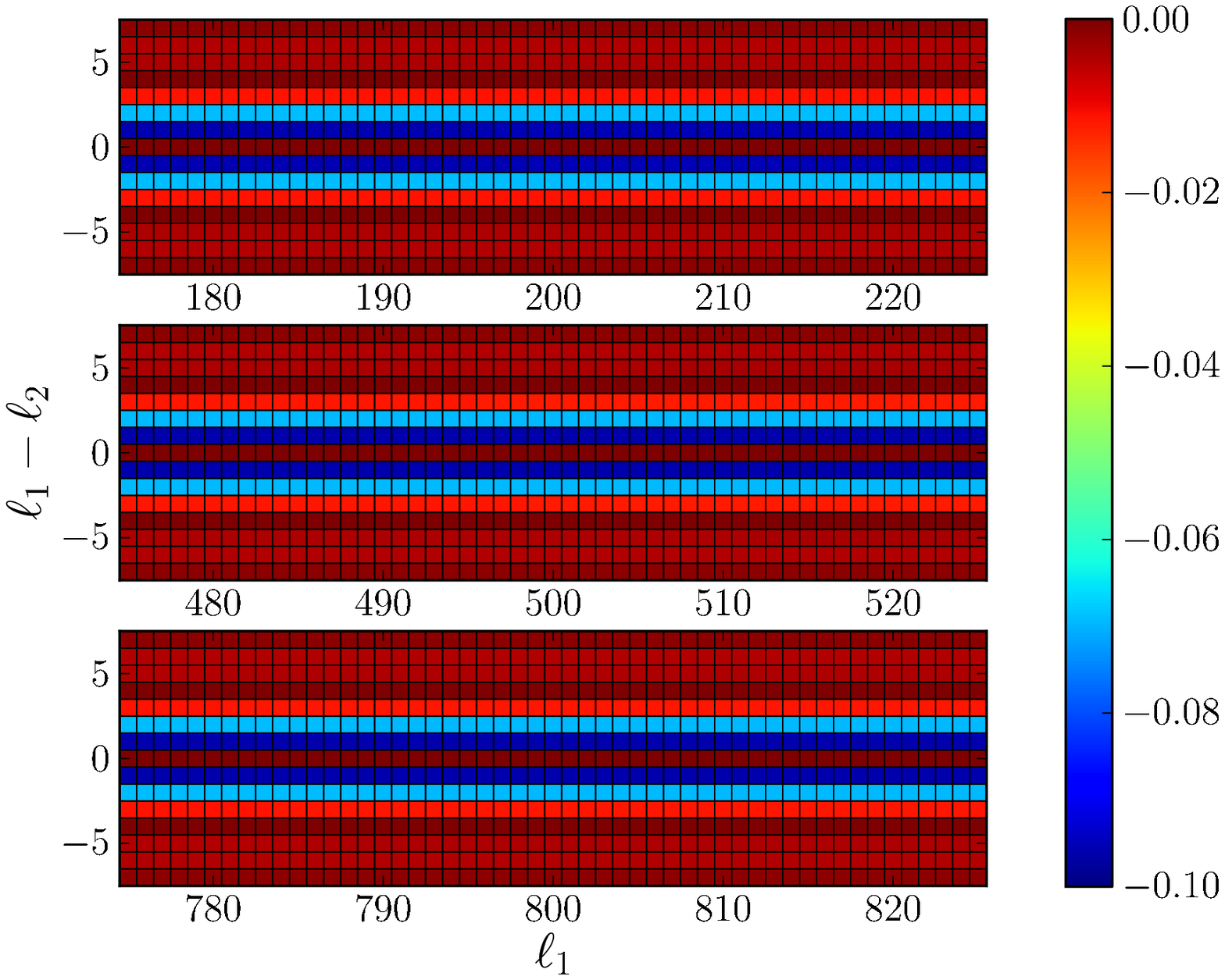}{fig:bench_correlation}
{The correlation structure between power spectrum coefficients is
  complicated. \emph{Left panel:} We plot the inverse of the Fisher
  matrix, i.e.\ the covariance matrix of the experiment. For
  illustrative purposes, we inverted the x and y axes. \emph{Right
    panel:} Patches around the diagonal elements of the normalized
  correlation matrix at low ($\ell = 200$, \emph{upper row}),
  intermediate ($\ell = 500$, \emph{middle row}), and high multipoles
  ($\ell = 800$, \emph{lower row}) reveal significant negative
  correlations between neighboring power spectrum coefficients at the
  10~\% level.}

\threefig{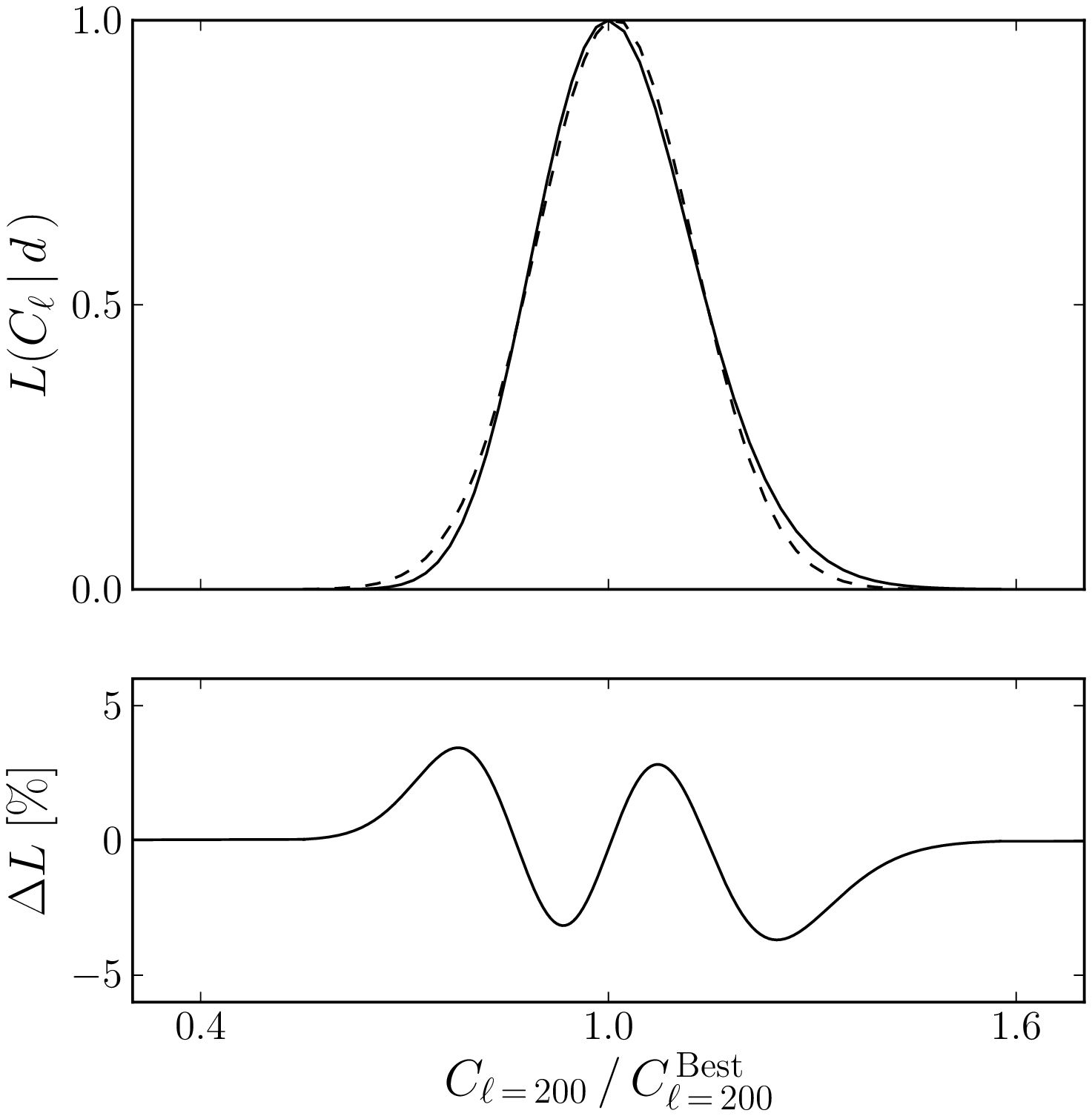}{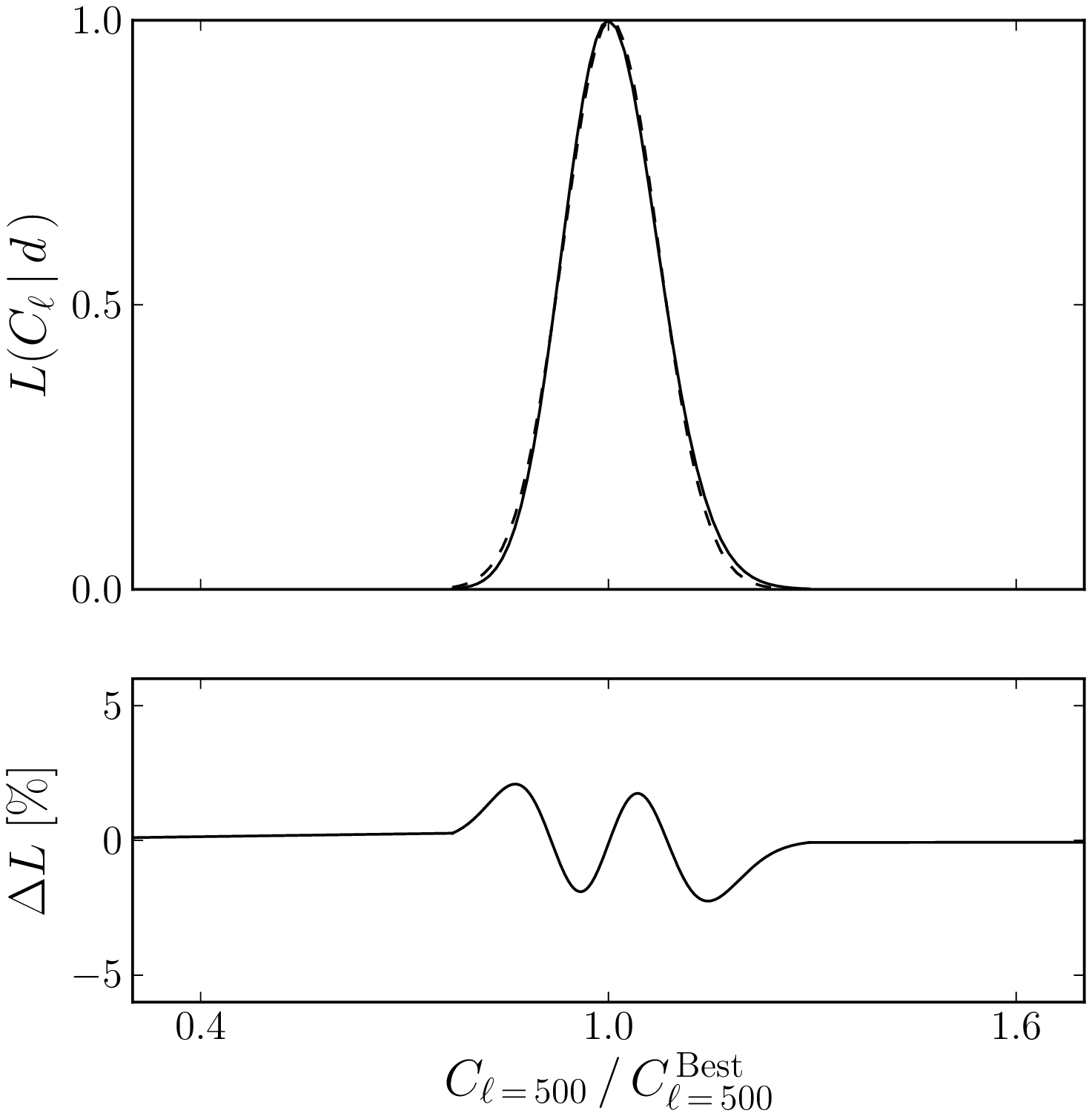}
{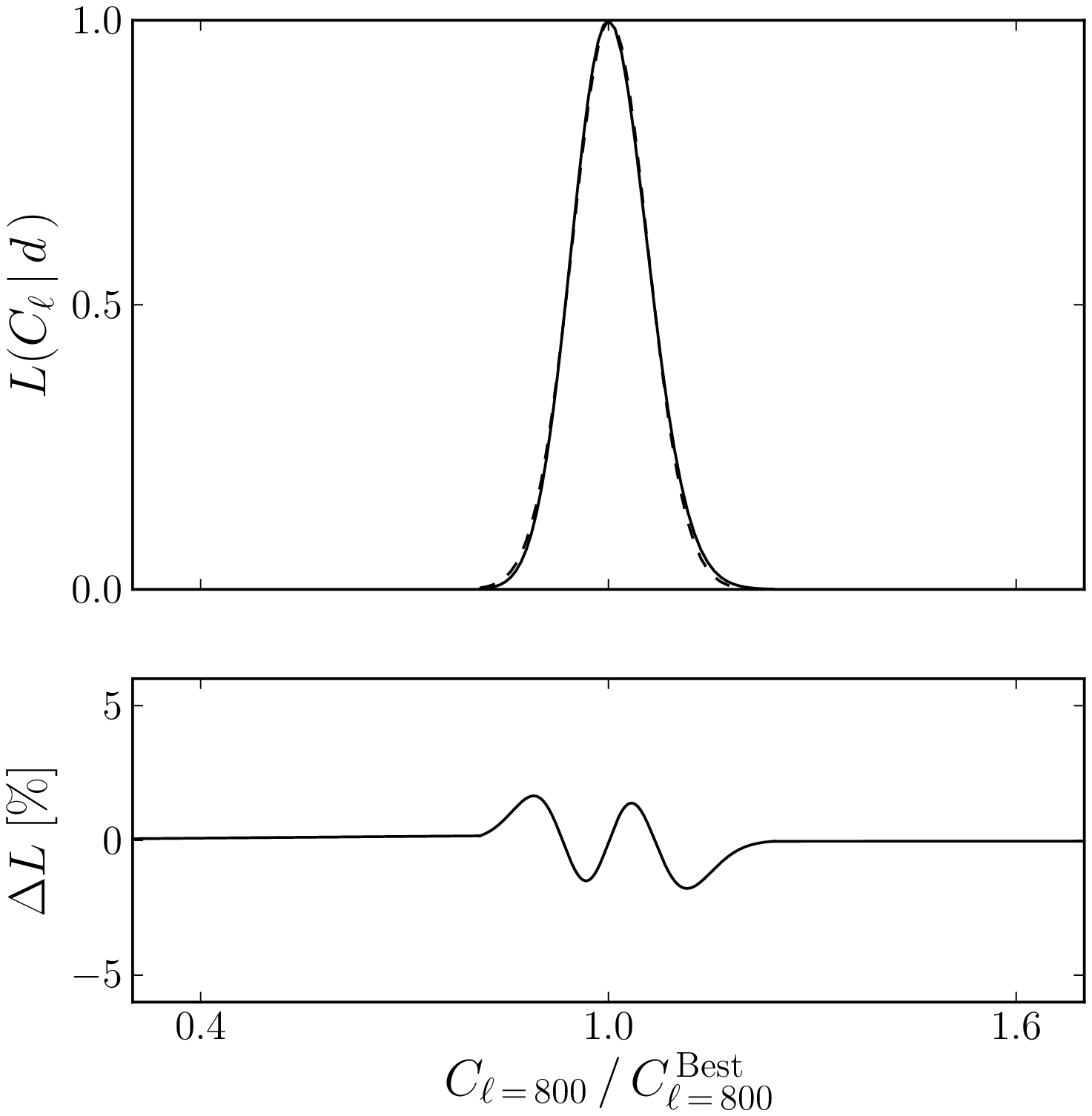}{fig:bench_like}
{The likelihood is non-Gaussian. \emph{Upper row:} Varying the power
  spectrum coefficients for different multipole values, we plot slices
  of the likelihood function (\emph{solid lines}) at low ($\ell =
  200$, \emph{left panel}), intermediate ($\ell = 500$, \emph{middle
    panel}), and high multipoles ($\ell = 800$, \emph{right
    panel}). We also show the best fitting Gaussian distributions
  (\emph{dashed lines}). \emph{Lower row:} The absolute residual
  between the likelihood function and its Gaussian fit is of the order
  of several percent.}

\subsection{The effect of asymmetric beams}

At some level, all CMB experiments suffer from beam imperfections,
depending on detector design and focal plane geometry. For the low
frequency instrument aboard the Planck satellite, for example, the
beam ellipticity can reach 40~\% \citep{2011A&A...536A...3M}. We now
study whether asymmetric beams by themselves pose a problem in a sense
that they lead to a loss of information or induce additional
correlations between power spectrum coefficients. Again, we study the
impact of asymmetric beams on the information content that could
ultimately be obtained after an optimal analysis of the data.

To this end, we repeated the full likelihood analysis on a data set
generated with the simulation parameters as above, except for the fact
that we now included an asymmetric Gaussian beam with an axes ratio of
$\sigma_{\mathrm{x}}/\sigma_{\mathrm{y}} = 2:1$. We kept the geometric
mean of the FWHM along the major and minor axis fixed at $12
\arcmin$. In \fig{fig:beam_fisher}, we compare a slice through the
covariance matrix of the simulation against our benchmark model. The
relative difference of the entries close to the dominating diagonal
elements is smaller than 1~\%. We conclude that an asymmetric beam per
se, if accounted for exactly in the analysis, neither degrades the
information content of the experiment, nor complicates the correlation
structure of the power spectrum coefficients.

\onefig{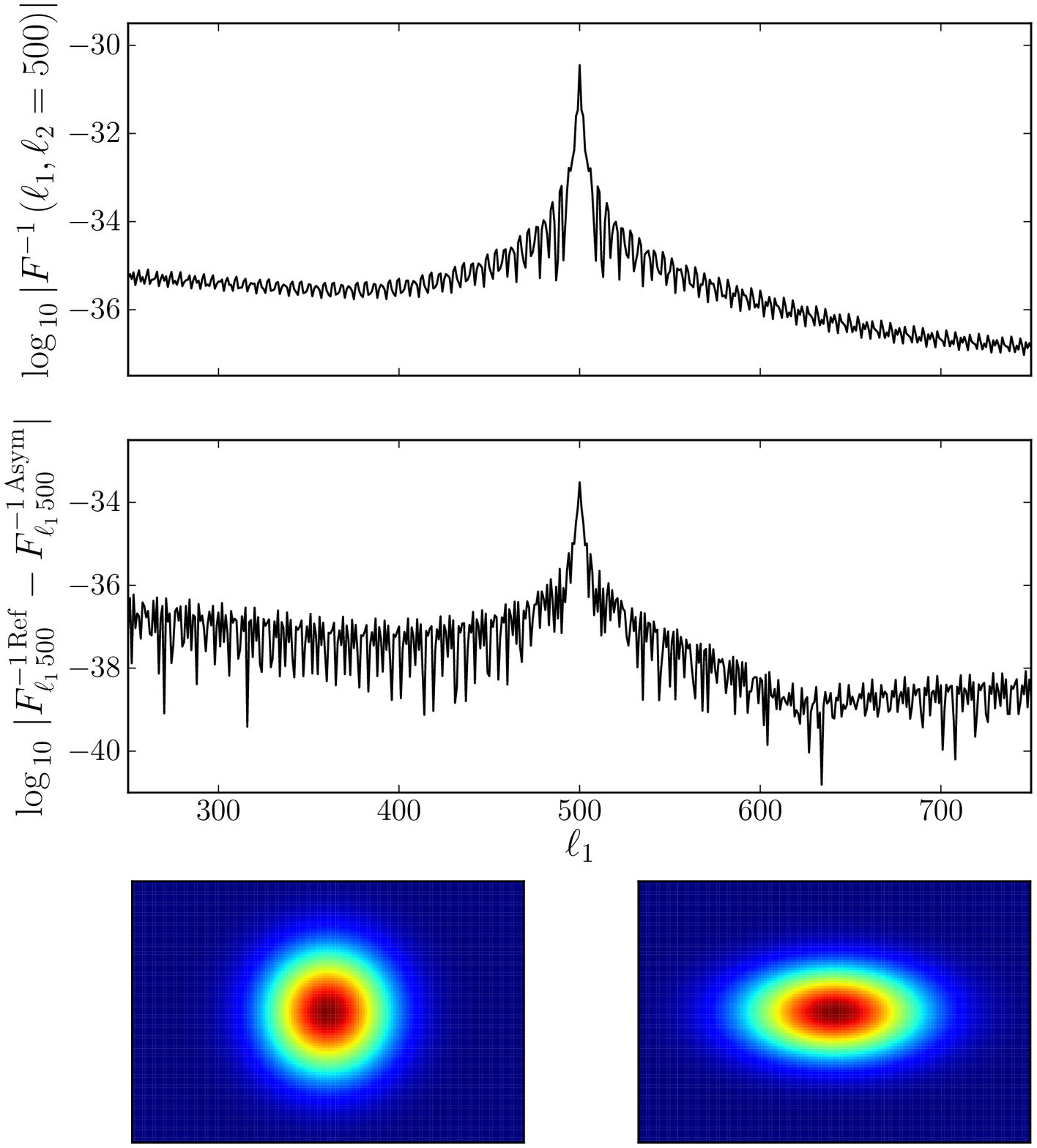}{fig:beam_fisher}
{Asymmetric beams do not destroy information. \emph{Upper panel:} For
  a multipole moment of $\ell = 500$, we show a slice through the
  covariance matrix for the reference simulation with symmetric
  beam. \emph{Middle panel:} The relative difference between the
  covariance matrices calculated with symmetric beam ($F^{-1 \,
    \mathrm{Ref}}$) and strongly asymmetric beam ($F^{-1 \,
    \mathrm{Asym}}$) is only of the order 1~\%. \emph{Lower
    panel:} Beam profiles used in the comparison.}

\subsection{The effect of the noise properties}

We now characterize the impact of different noise properties on the
likelihood function. To be more precise, we first vary the overall
noise level without modifying its power spectrum, and then analyze the
effect of correlated noise. The data sets used for this study are
visualized in \fig{fig:noise_torus}.

\subsubsection{Low S/N likelihood}

For the analysis of the effect of noise on CMB power spectrum
estimation, we increased the noise power spectrum amplitude in
\eq{eq:noise} to $\sigma = 200$~mK. In this setup, we obtained a S/N
of unity at a multipole $\ell \approx 700$. As expected, the diagonal
elements of the inverse Fisher matrix reflect the increase in noise
(\fig{fig:low_snr_fisher}). We also find the correlation structure of
the Fisher matrix change qualitatively at high multipoles, it becomes
stronger and more complex. In particular, the negative correlation
between a power spectrum coefficient $C_{\ell}$ and $C_{\ell \pm 5}$
increases in strength. The full correlation matrix, plotted on
logarithmic scale to highlight the off-diagonal entries, is depicted
in \fig{fig:full_cm_low_snr_uncor}. Becoming more and more relevant
with the decreasing S/N towards higher $\ell$, the inhomogeneously
distributed noise induces additional long-ranging correlations between
power spectrum coefficients at the sub-percent level. We show
likelihood slices in \fig{fig:low_snr_like}. When compared to the high
S/N case in \fig{fig:bench_like}, the analysis reveals a widened
distribution at high multipoles. Besides that, the non-Gaussian
features of the likelihood function remain basically unchanged. Only a
further decrease in the S/N starts to diminish the deviation from a
Gaussian shape. In that case, however, the power spectrum coefficients
have too large an error bar for them to play a substantial role in the
estimation of cosmological parameters.

\onelargefig{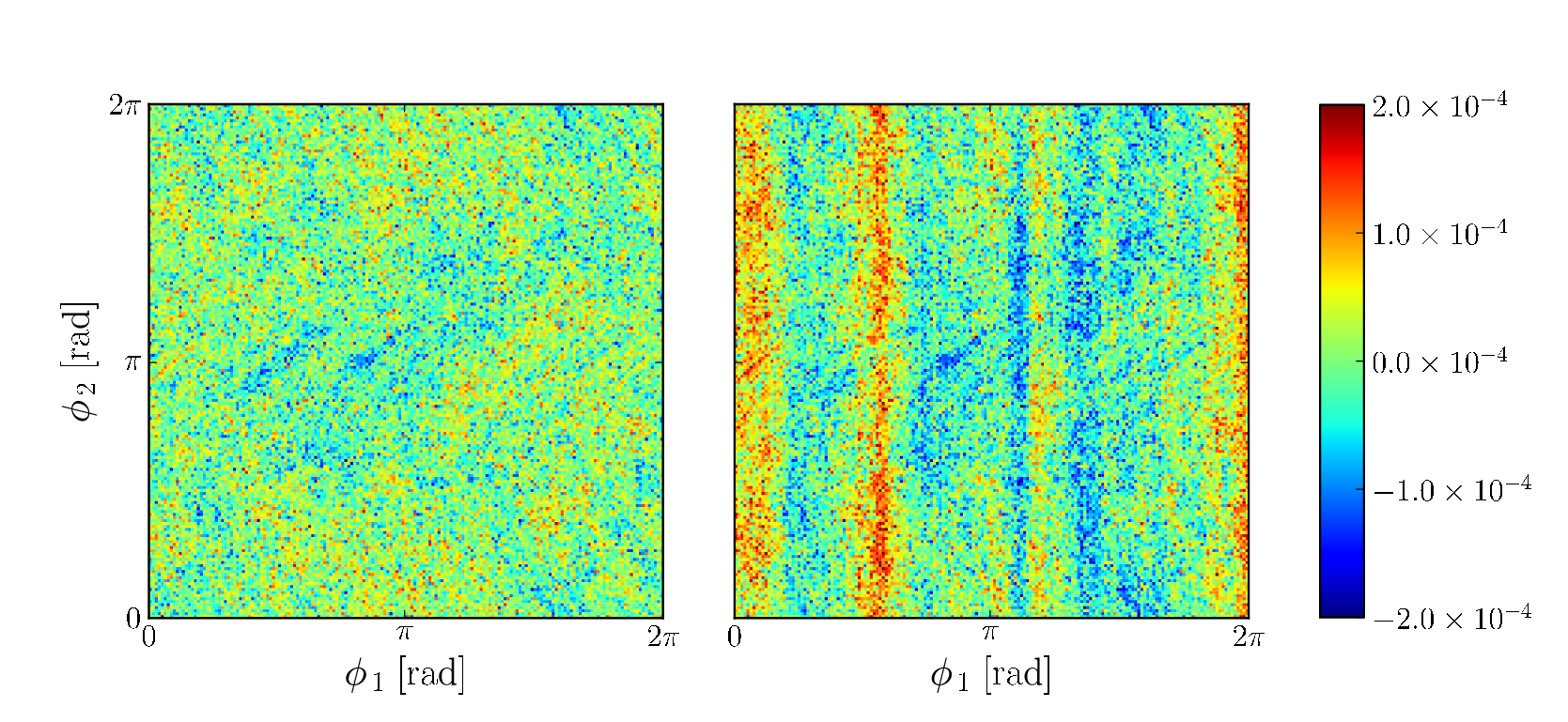}{fig:noise_torus}
{Data sets used to study the impact of noise. \emph{Left panel:} We
  show the data of the low S/N simulation with uncorrelated
  noise projected onto an unfolded torus. The scan on a single ring
  proceeds towards larger values of $\phi_{2}$, the ring number
  increases along the horizontal axis ($\phi_{1}$). \emph{Right
    panel:} Low S/N simulation, additionally containing
  $1/f^2$-type noise. Strong correlations extending well beyond
  individual rings are visible.}

\twofig{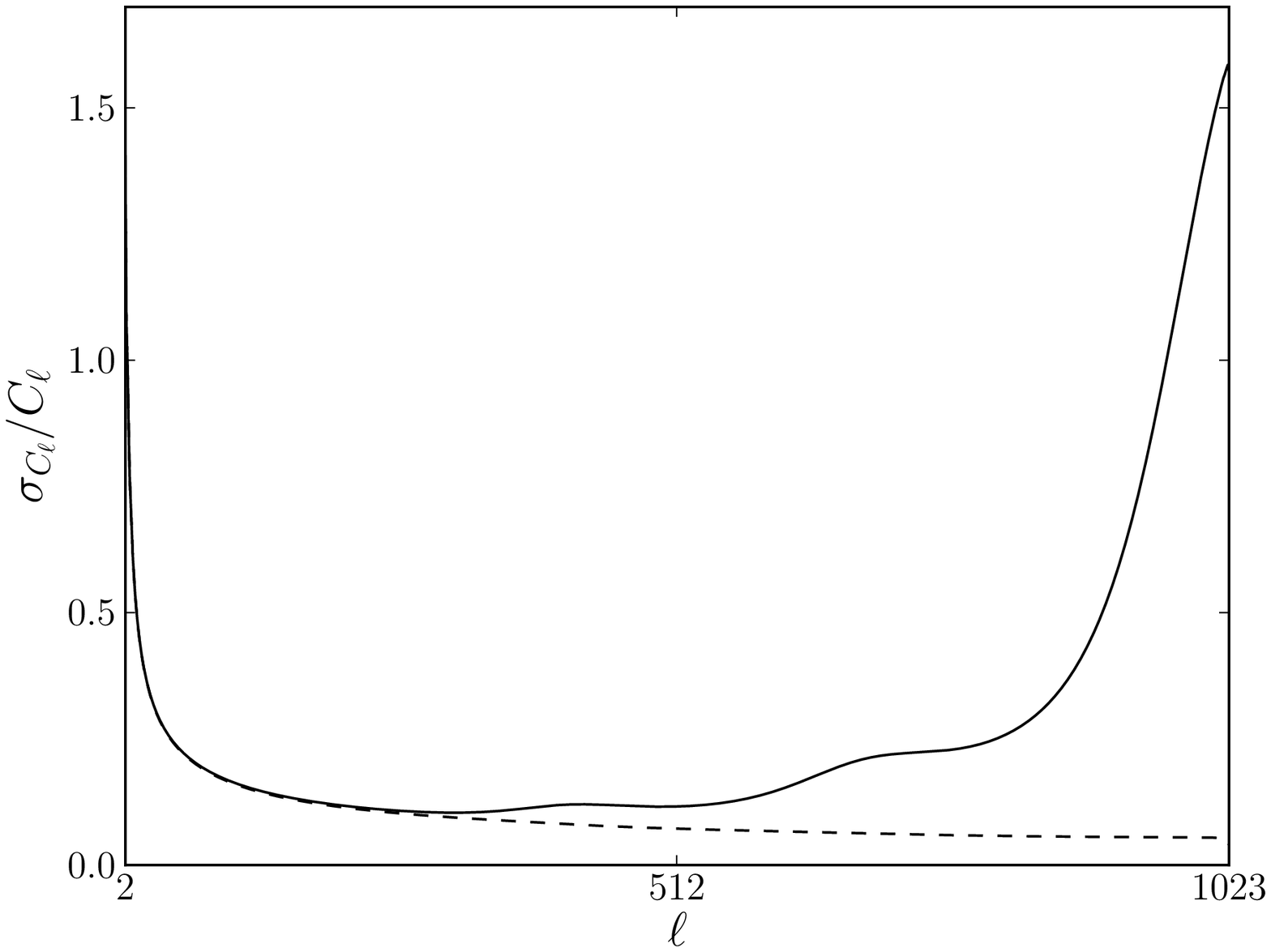}{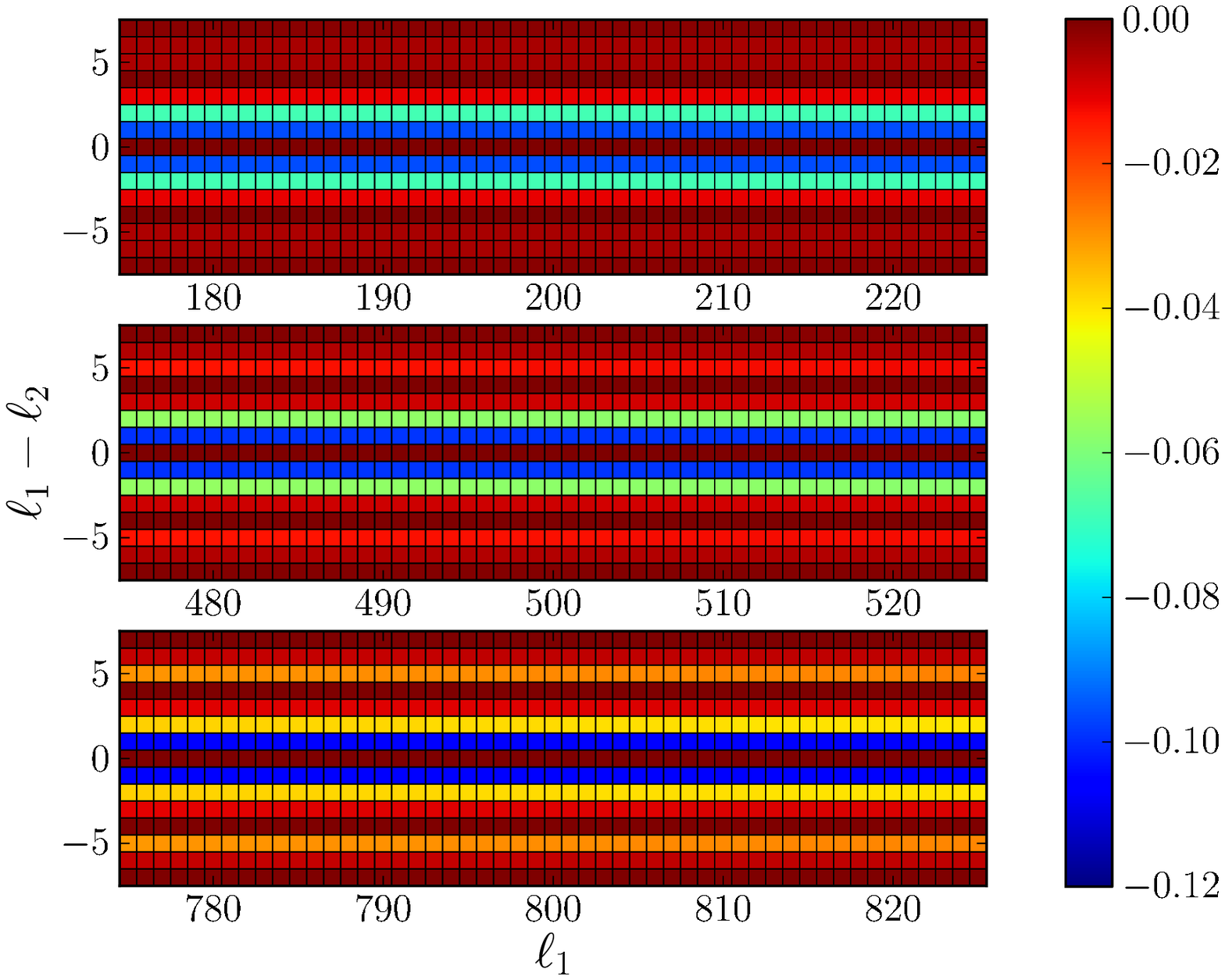}
{fig:low_snr_fisher}
{\emph{Left panel}: We compare the Fisher matrix based prediction of
  the 1-$\sigma$ error bars of the power spectrum coefficients between
  our benchmark (\emph{dashed line}) and the simulation with lower S/N
  and uncorrelated noise (\emph{solid line}). \emph{Right
    panel}: As a plot of the correlation matrix confirms, an increase
  in noise towards higher multipoles strengthens the amount of
  correlation and complicates the correlation structure
  (cf.\ \fig{fig:bench_correlation}).}

\onefig{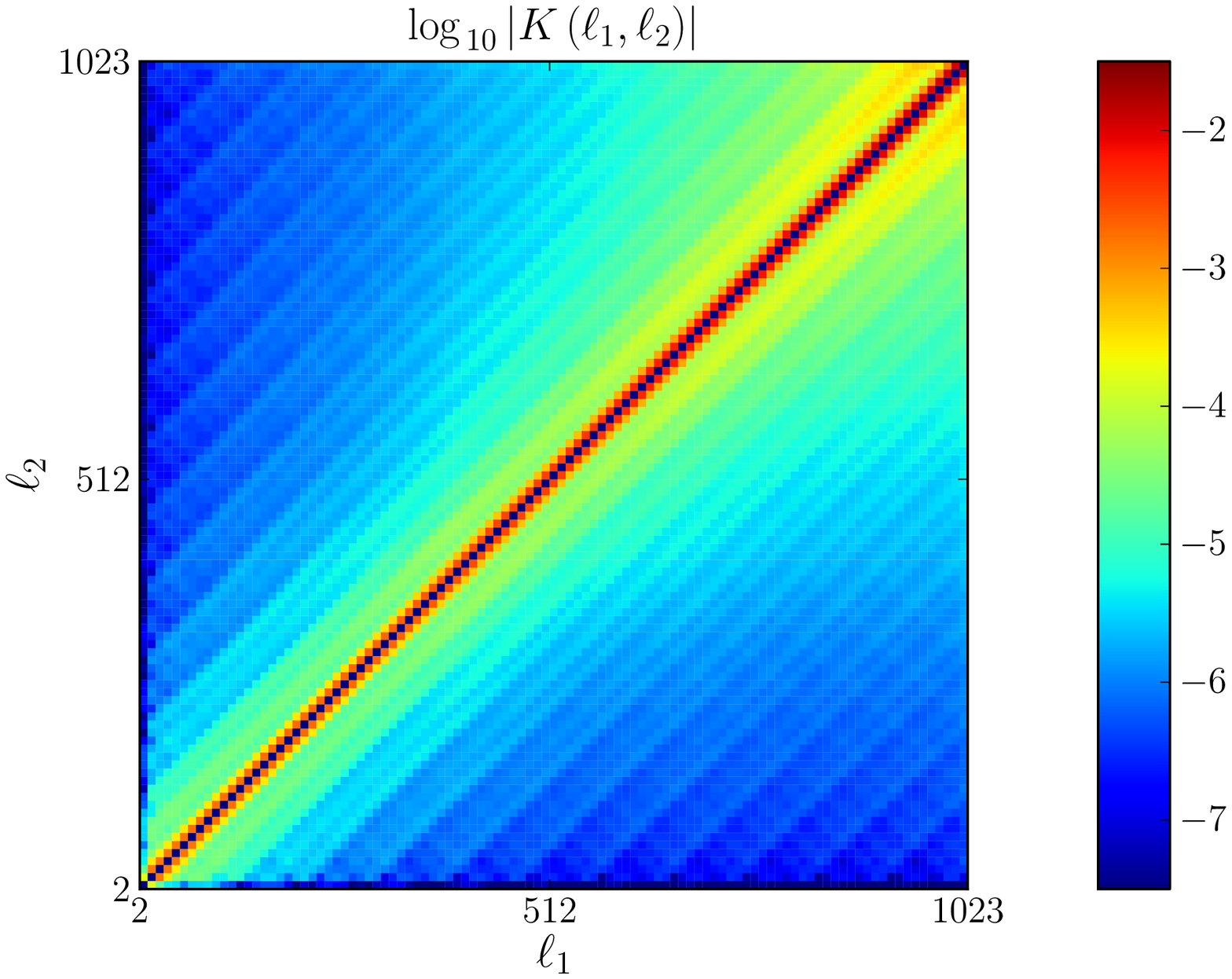}{fig:full_cm_low_snr_uncor}
{An increased level of noise, distributed inhomogenously over the sky,
  introduces additional long-range correlations between power spectrum
  coefficients at the sub-percent level.}

\threefig{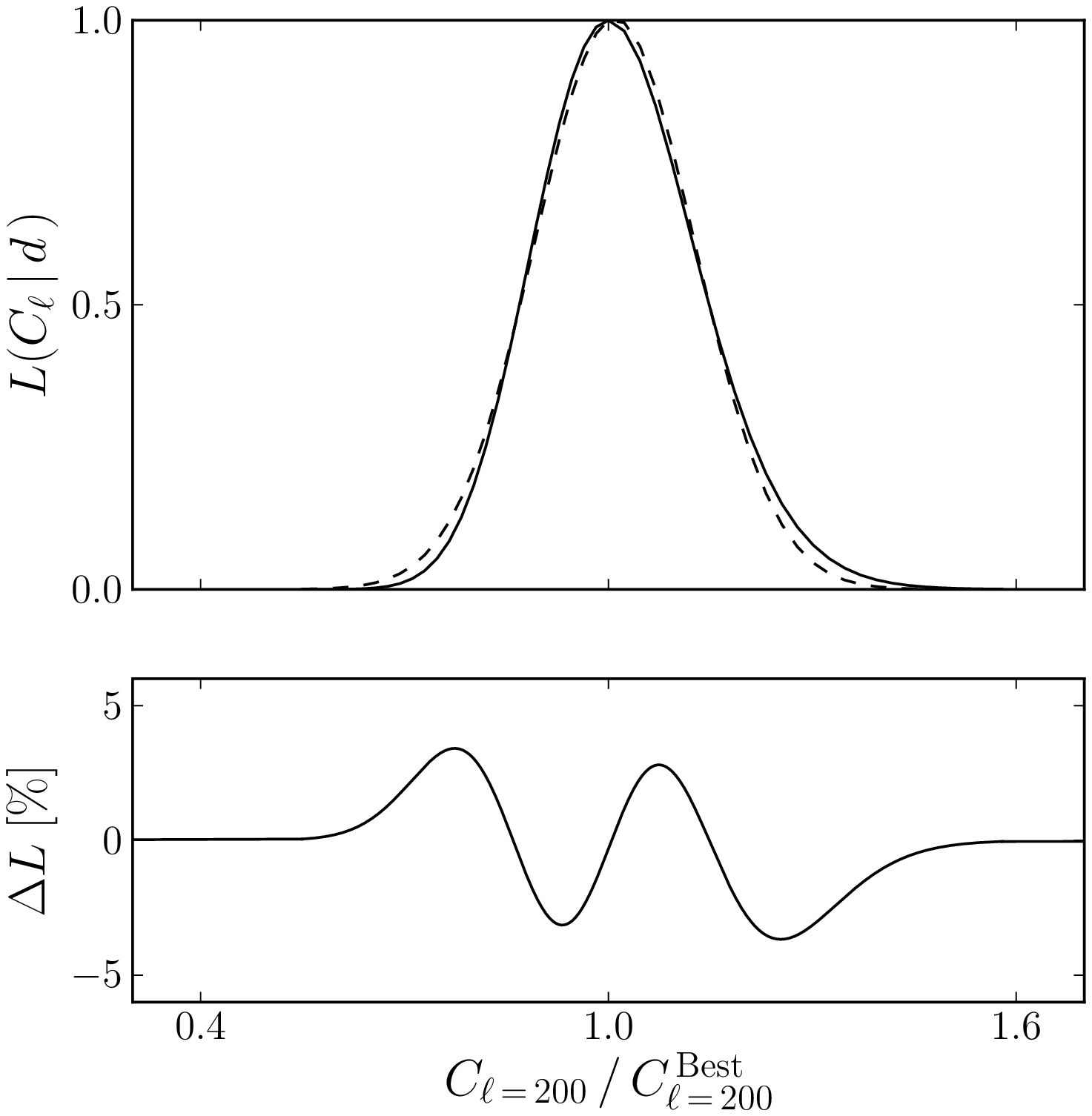}{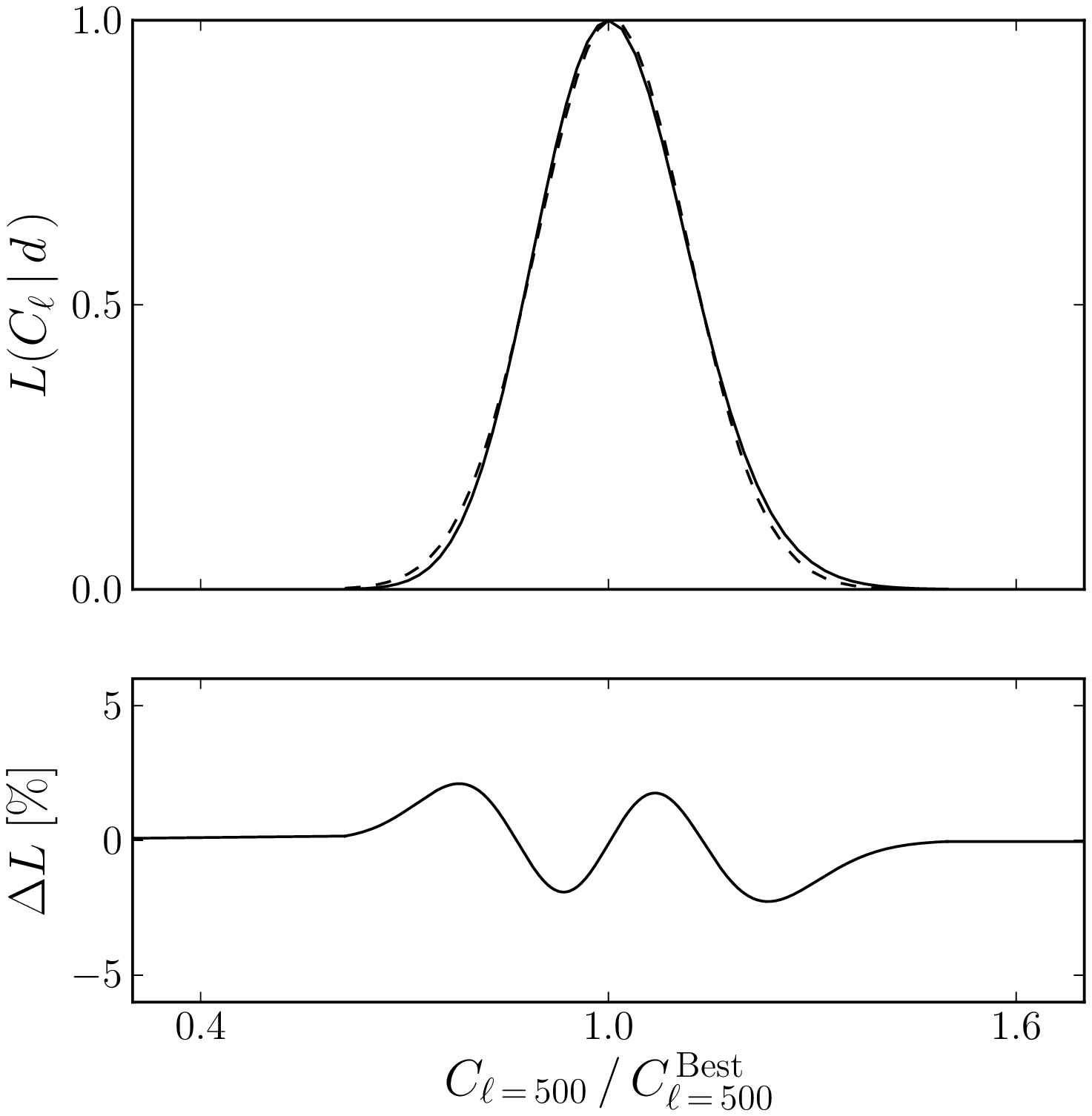}
{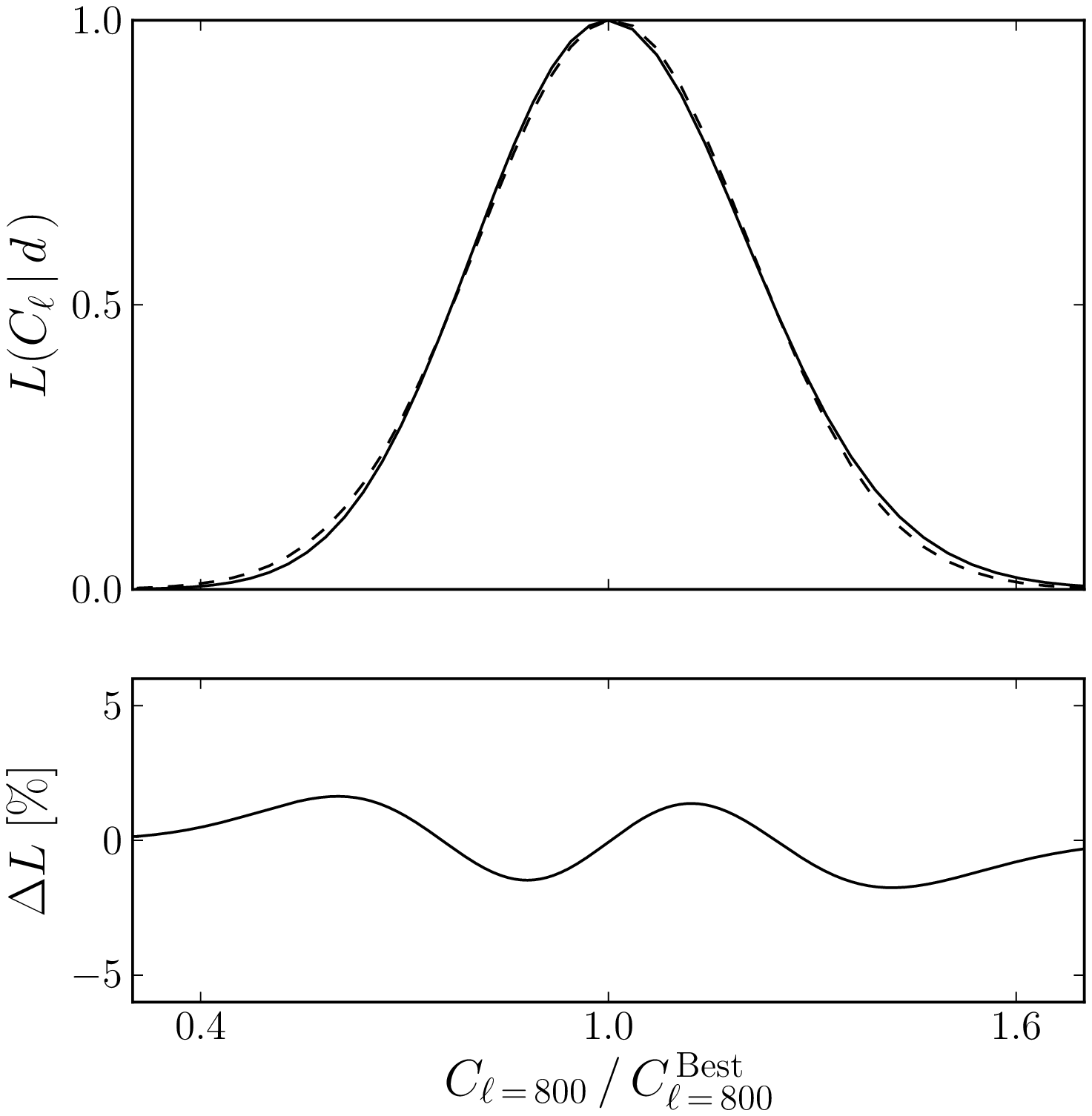}{fig:low_snr_like}
{Same as \fig{fig:bench_like}, but for a simulation with low
  S/N. Simply increasing the error bars, additional noise, most
  noticeable at the highest multipole moment, largely preserves the
  non-Gaussian features of the likelihood function.}

\subsubsection{Correlated noise}

Next, we study the effect of correlated noise. To this end, we
generate a simulation with parameters as follows. Choosing a noise
realization with $\alpha = 2$ in \eq{eq:noise}, we selected a knee
frequency of $k_{\mathrm{knee}} = 10^{-3}$ in terms of the Nyquist
frequency, and set the DC mode to zero. Red noise is often present in
experiments to some level, and it also affects the high frequency
instrument data of the Planck satellite mission
\citep{2011A&A...536A...4P}. Here, we kept the overall noise power
spectrum amplitude at $\sigma = 200$~mK, the same level as in the
previous scenario. Owing to the additional long-range correlations,
the noise variance per pixel in the data map increased considerably
(cf.\ \fig{fig:noise_torus}).

The analysis based on the Fisher matrix is shown in
\fig{fig:corr_noise_fisher}, where we plot the error bars and the
correlation matrix of the experiment. The impact of the red noise on
the likelihood function is negligible when it is accounted for exactly
in the analysis. The additional low-frequency noise contribution,
most relevant for large-scale perturbations, still falls within a
highly signal dominated regime. As a result, its effect on the power
spectrum estimation can be removed efficiently. Only for the smallest
multipole moments, $\ell \la 5$, we observe an increase in the error
bar by no more than 3~\%. Accordingly, the correlation structure
between power spectrum coefficients is virtually unaffected.

\twofig{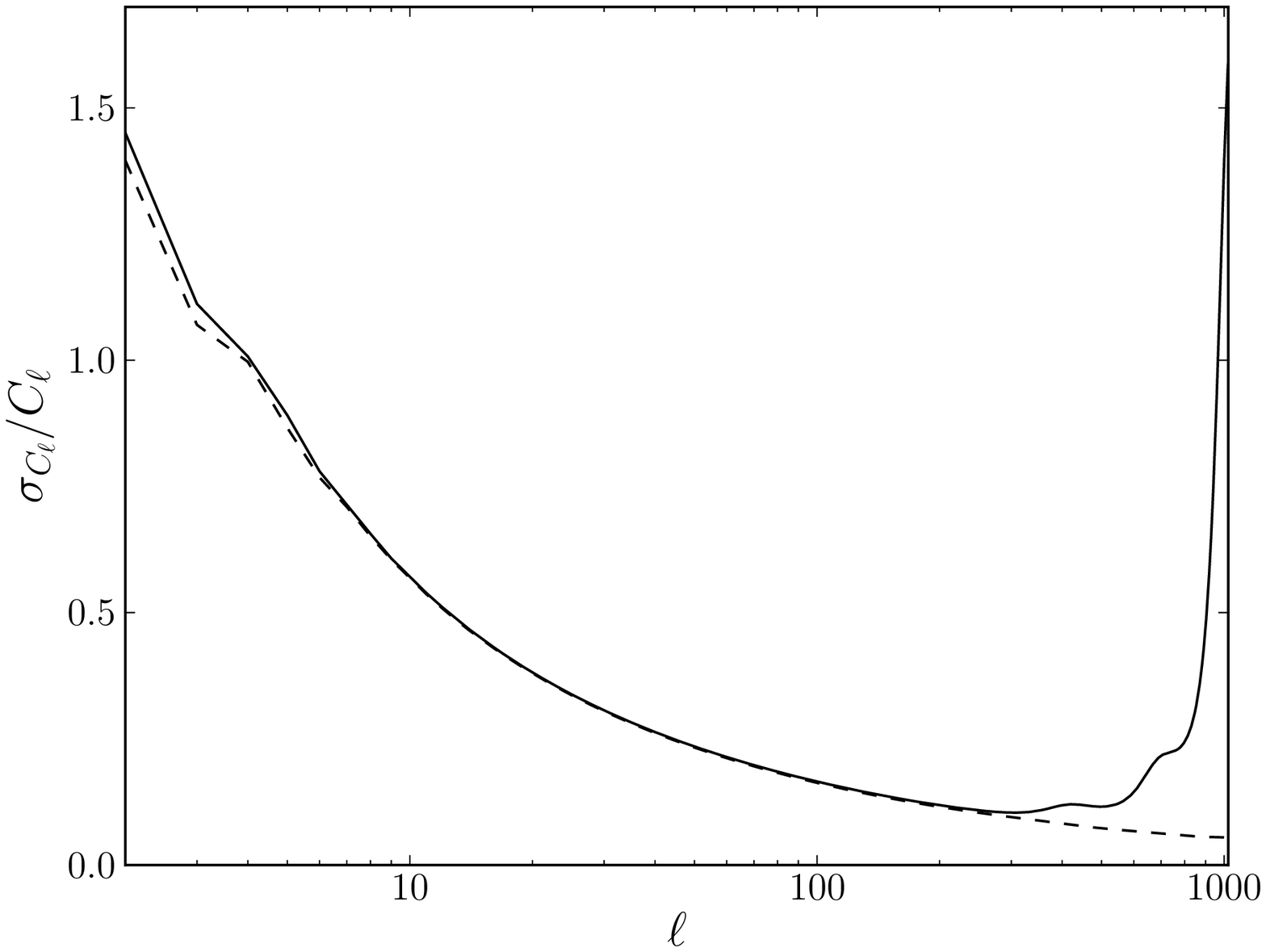}{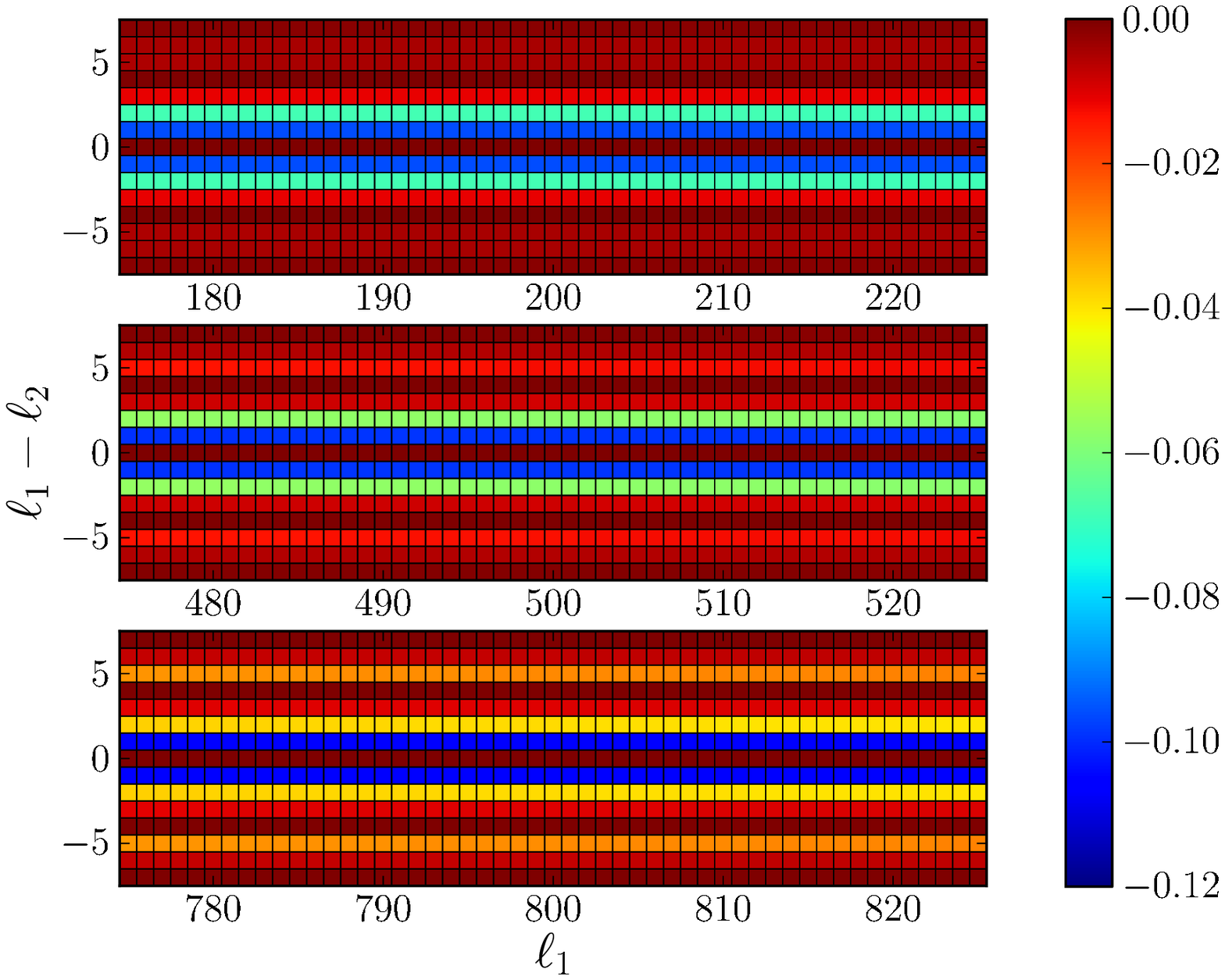}
{fig:corr_noise_fisher}
{Same as \fig{fig:low_snr_fisher}, but for the simulation containing
  noise with additional large-scale correlations. The error bars
  remain at the same level. Only at the largest scales, $\ell < 5$, a
  mild loss of information is apparent. Changes in the normalized
  correlation matrix are negligible.}

\subsection{The effect of partial sky coverage}

Excluding large parts of the sky from the analysis has the potential
to severely complicate power spectrum estimations. This problem roots
in the mathematical properties of the spherical harmonics. Restricted
to a subdomain of the sphere, the functions do no longer fulfill an
orthonormality relation. As a result, correlations between the CMB
power spectrum coefficients are induced.

All ground based CMB experiments observe only a fraction of the sky
\citep[e.g.,][]{2003NewAR..47..939K, 2004SPIE.5498...11R}. But even
for satellite-borne experiments, the contamination by secondary
sources necessitates to mask large sky regions in favor of an unbiased
CMB power spectrum estimation \citep[e.g.,][]{2011ApJS..192...14J}. To
further study the effect of restricting the observations to a small
part of the sky, we analyzed a simulation with a reduced coverage of
$f_{\mathrm{sky}} = 40~\%$. We chose the simulation parameters of our
benchmark data set, except for the opening angle of the scanning
rings, which we reduced to $\theta_{\mathrm{o}} = 25 \degr$. As we
still sampled the data with the same number of nodes ($N_{\mathrm{p}}
= N_{\mathrm{r}} = 2048$), we slightly adapted the noise amplitude to
the finer grid to obtain the same S/N in the power spectrum, $\sigma =
\sqrt{60~\%/40~\%} \cdot 10~\mathrm{mK} \approx 12.2$~mK. In
\fig{fig:fsky_fisher}, we visualize the normalized correlation
matrix. A comparison to the right hand panel of
\fig{fig:bench_correlation} shows an increase in the amplitude of
correlation by about a factor of two. In addition, an even/odd
asymmetry becomes visible as the power spectrum coefficients
$C_{\ell}$ are now negatively correlated with $C_{\ell \pm 3}$ at the
12~\% level, whereas the amount of negative correlation for $C_{\ell
  \pm 2}$ decreases to 5~\%. We plot slices of the likelihood in
\fig{fig:fsky_like}. The loss of information associated with a
reduction of sky coverage leads to widened distributions, most
noticeable at large angular scales. Interestingly, the non-Gaussian
pattern in the likelihood function are remarkably robust and remain at
about the same level.

\onefig{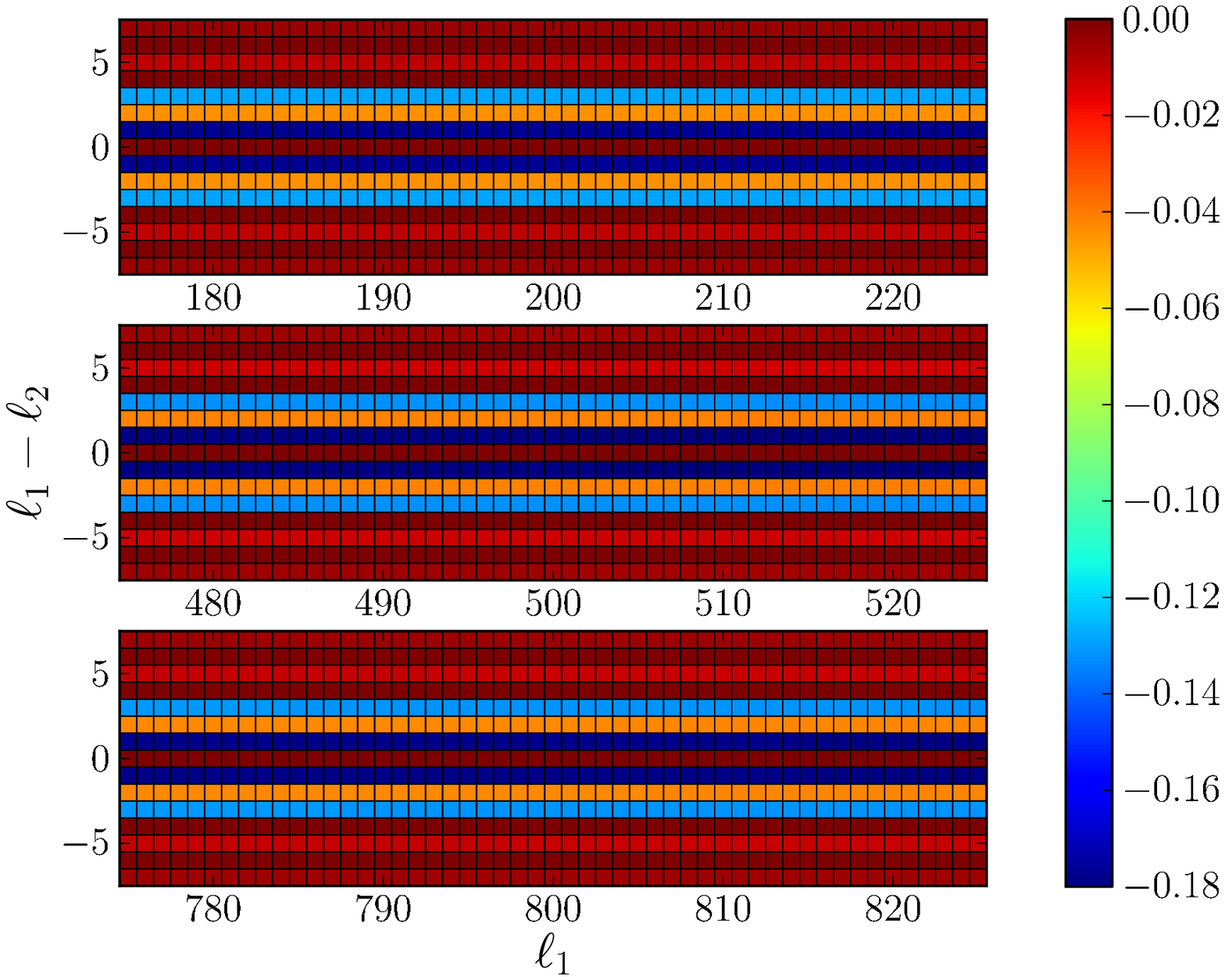}{fig:fsky_fisher}
{A small sky coverage induces complicated correlations. We plot the
  normalized correlation matrix for a simulation covering
  $f_{\mathrm{sky}} = 40~\%$ of the sky. A larger mask increases the
  correlation in strength and range
  (cf.\ \fig{fig:bench_correlation}).}

\threefig{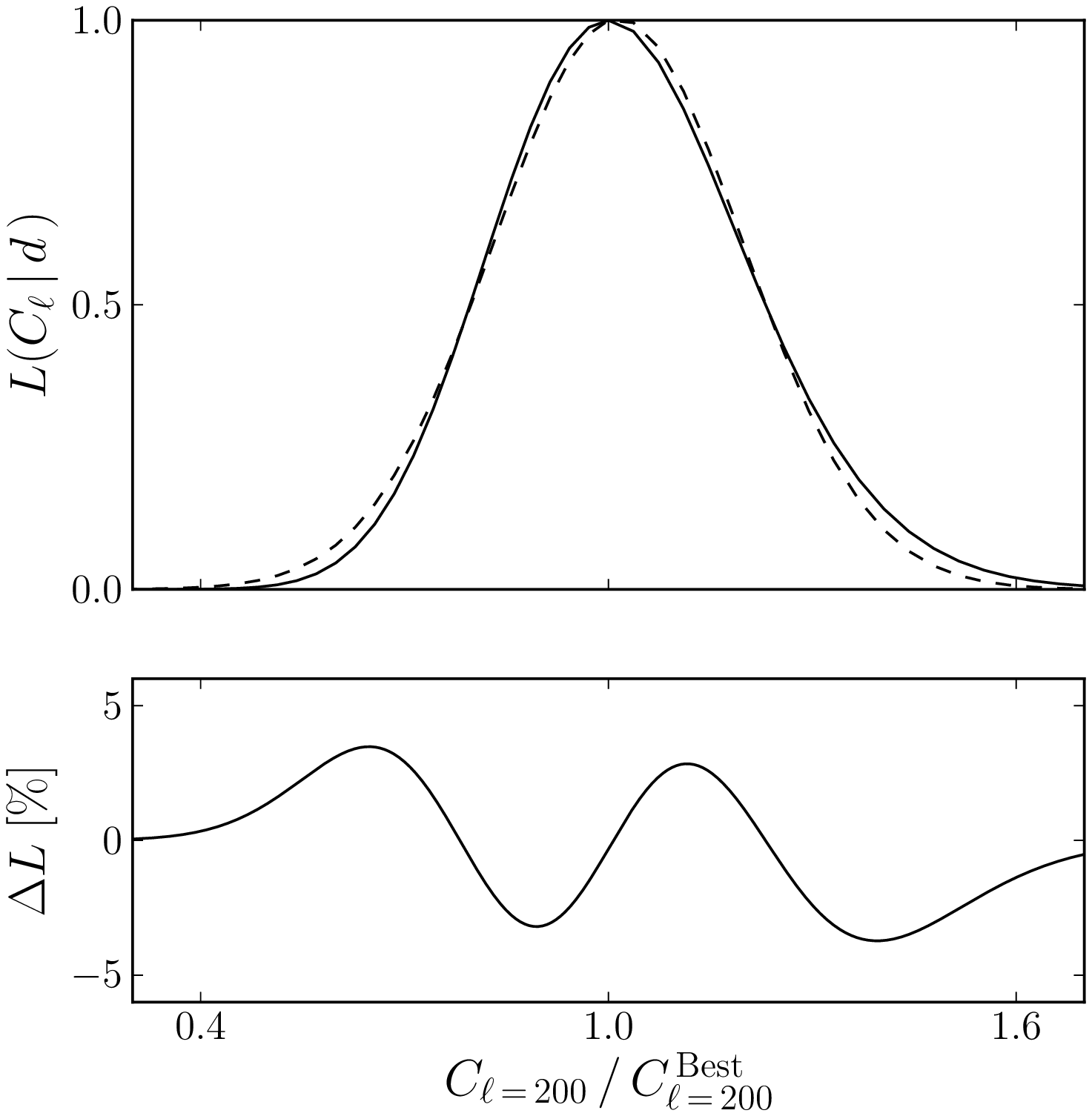}{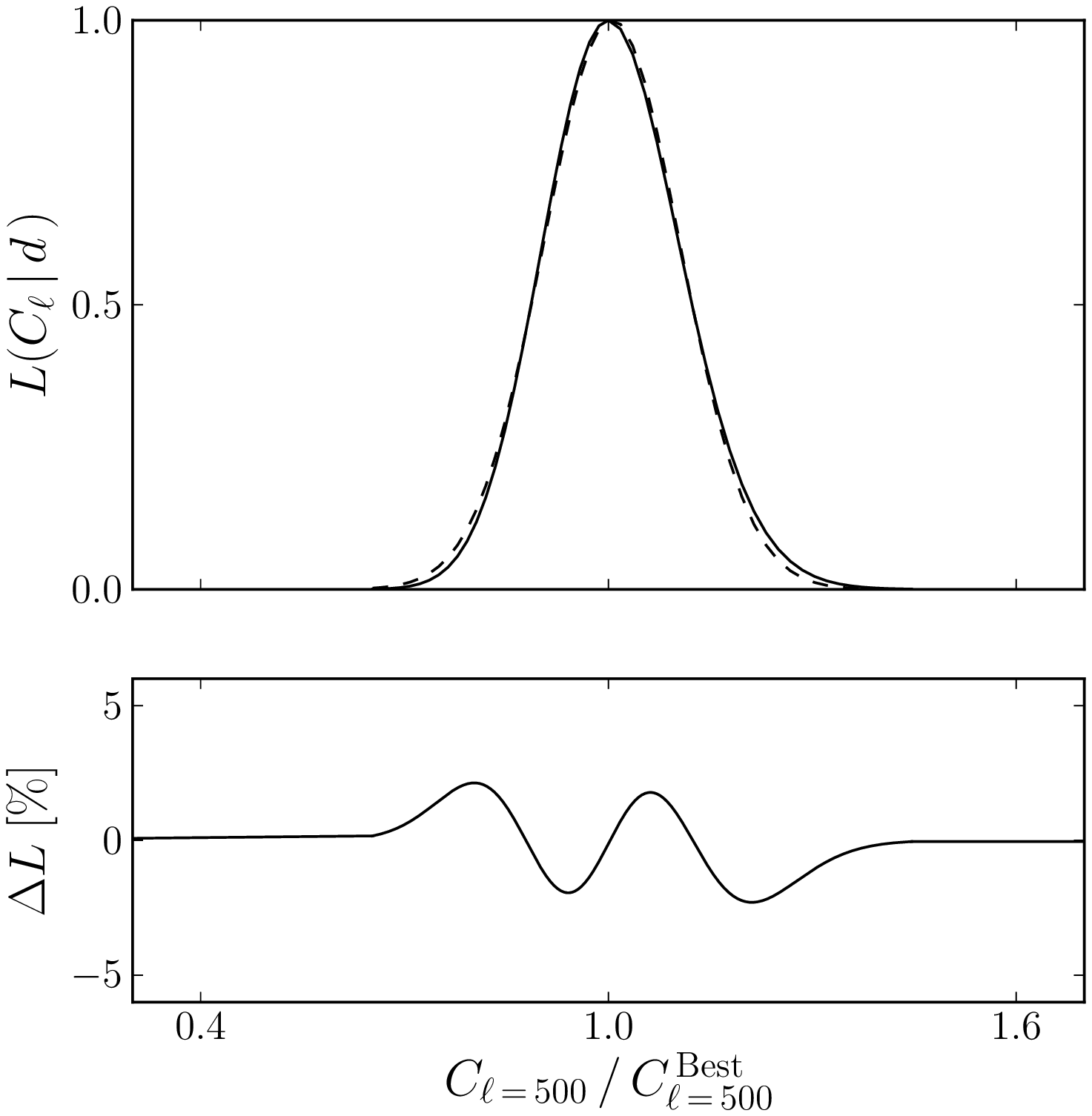}
{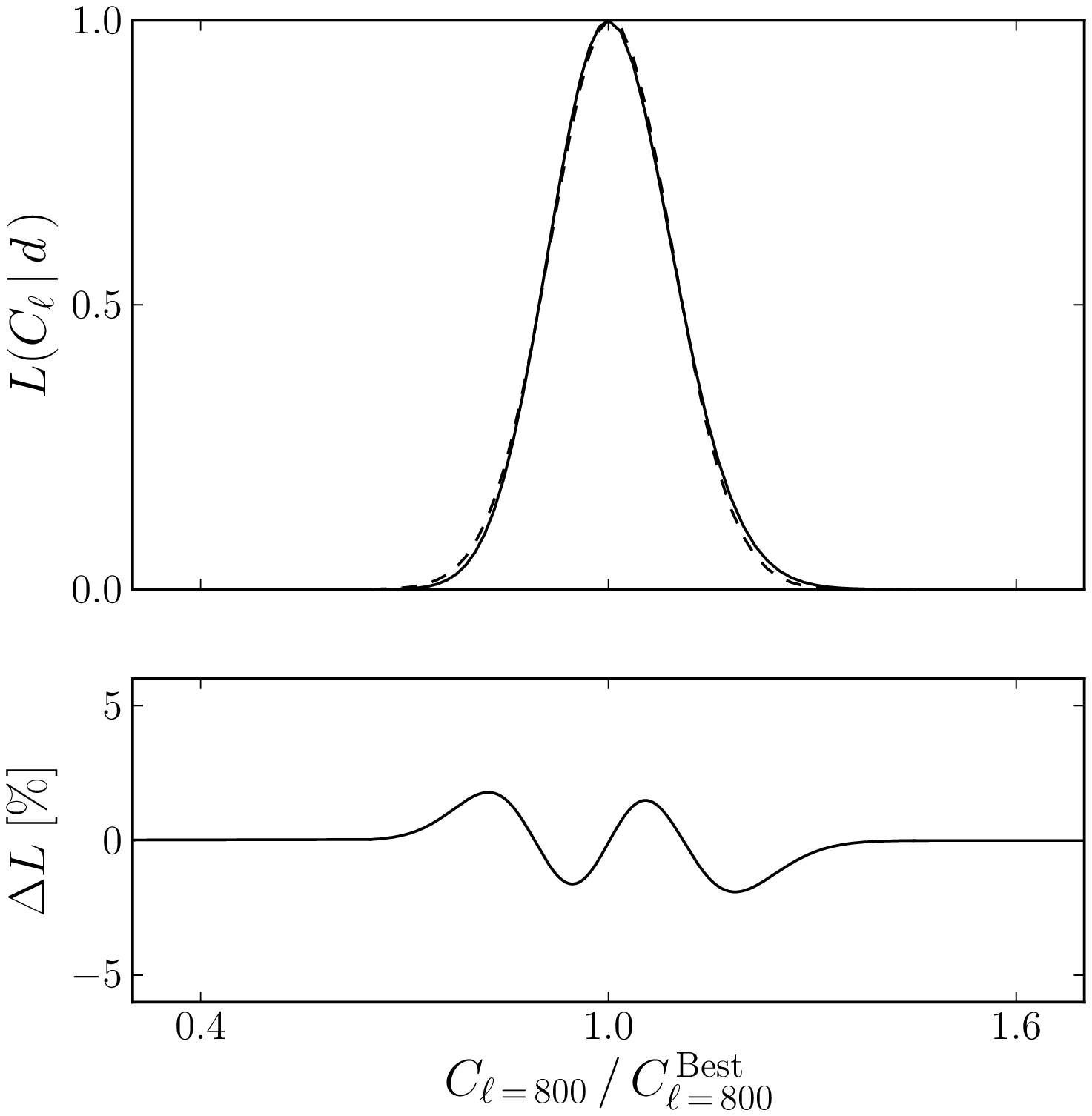}{fig:fsky_like}
{Same as \fig{fig:bench_like}, but for a simulation with larger
  mask. Lowering the sky coverage from 60~\% to 40~\% leads to less
  stringent constraints on the power spectrum coefficients, but has a
  negligible effect on the amount of deviation of the likelihood
  function from a Gaussian distribution.}

\subsection{The effect of binning}

As established in the previous paragraph, a large galactic cut induces
significant correlations between neighboring power spectrum
coefficients. Furthermore, the likelihood function is non-Gaussian to
a noticeable extent. As the theoretical power spectrum is a smooth
function of only a few cosmological parameters, combining the observed
power spectrum coefficients into bins is a common practice in
parameter estimation. This approach reduces scatter in the data points
and has the potential to simplify the correlation structure. Here, we
analyze the effect of binning on the likelihood function.

We chose a bin width of $\Delta \ell = 10$ and repeated the analysis
of our reference simulation. In \fig{fig:bin_fisher}, we plot the
correlation structure of the likelihood function. Neighboring power
spectrum bins are still negatively correlated, although at a smaller
amount of about 5~\%. The bin after next, however, is almost
completely decorrelated. Furthermore, the shape of the likelihood
function, shown in \fig{fig:bin_like}, can be much better described by
a Gaussian distribution, where the residuals of a fit have decreased
to a 1 - 2~\% level. Obviously, binning has the desirable side effects
to simplify the shape and correlation structure of the likelihood
function.

\onefig{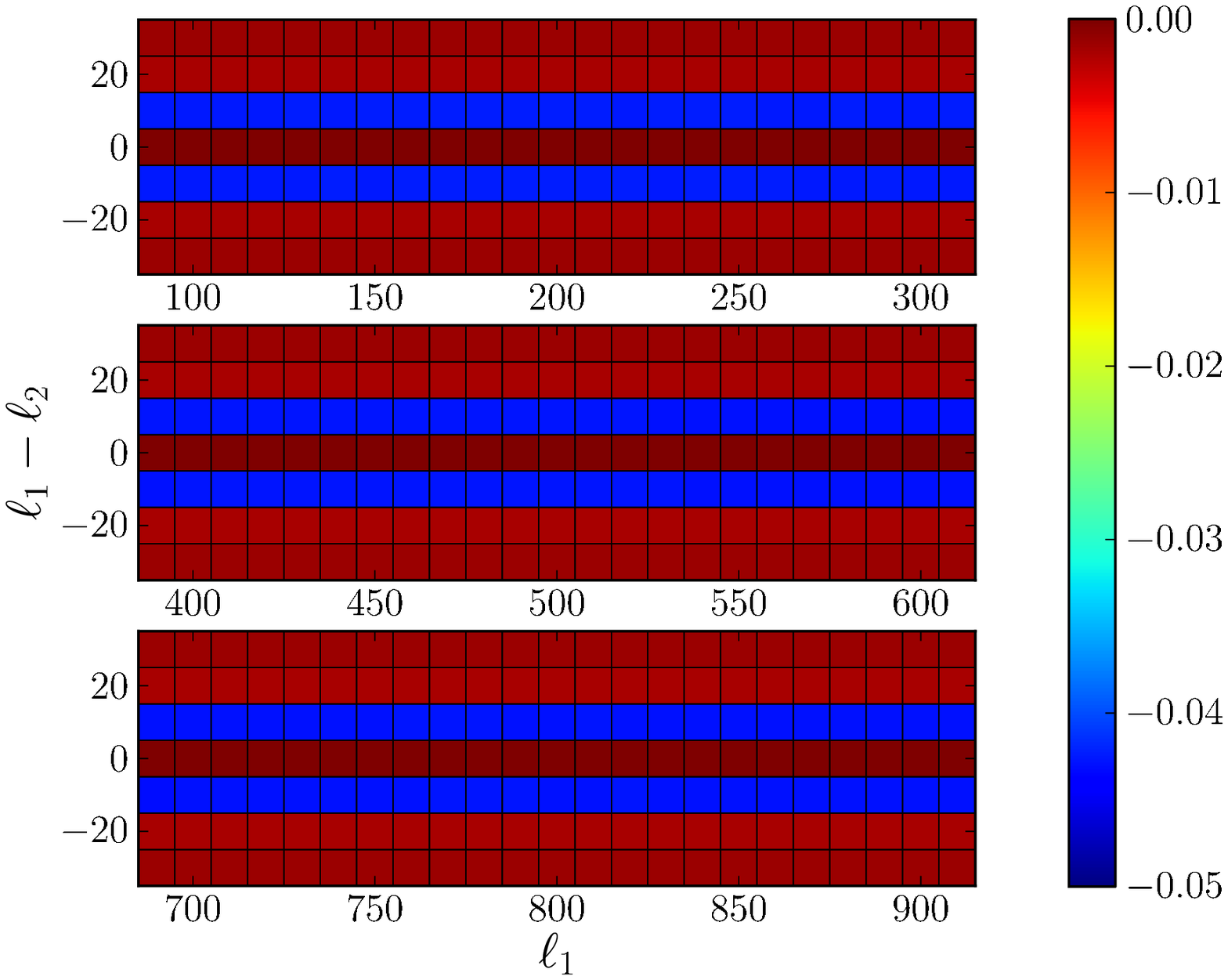}{fig:bin_fisher}
{Binning simplifies the correlation structure. We show elements of the
  correlation matrix for a binned power spectrum with a width of
  $\Delta \ell = 10$. Directly neighboring bins are still negatively
  correlated to a noticeable extent
  (cf.\ \fig{fig:bench_correlation}).}

\threefig{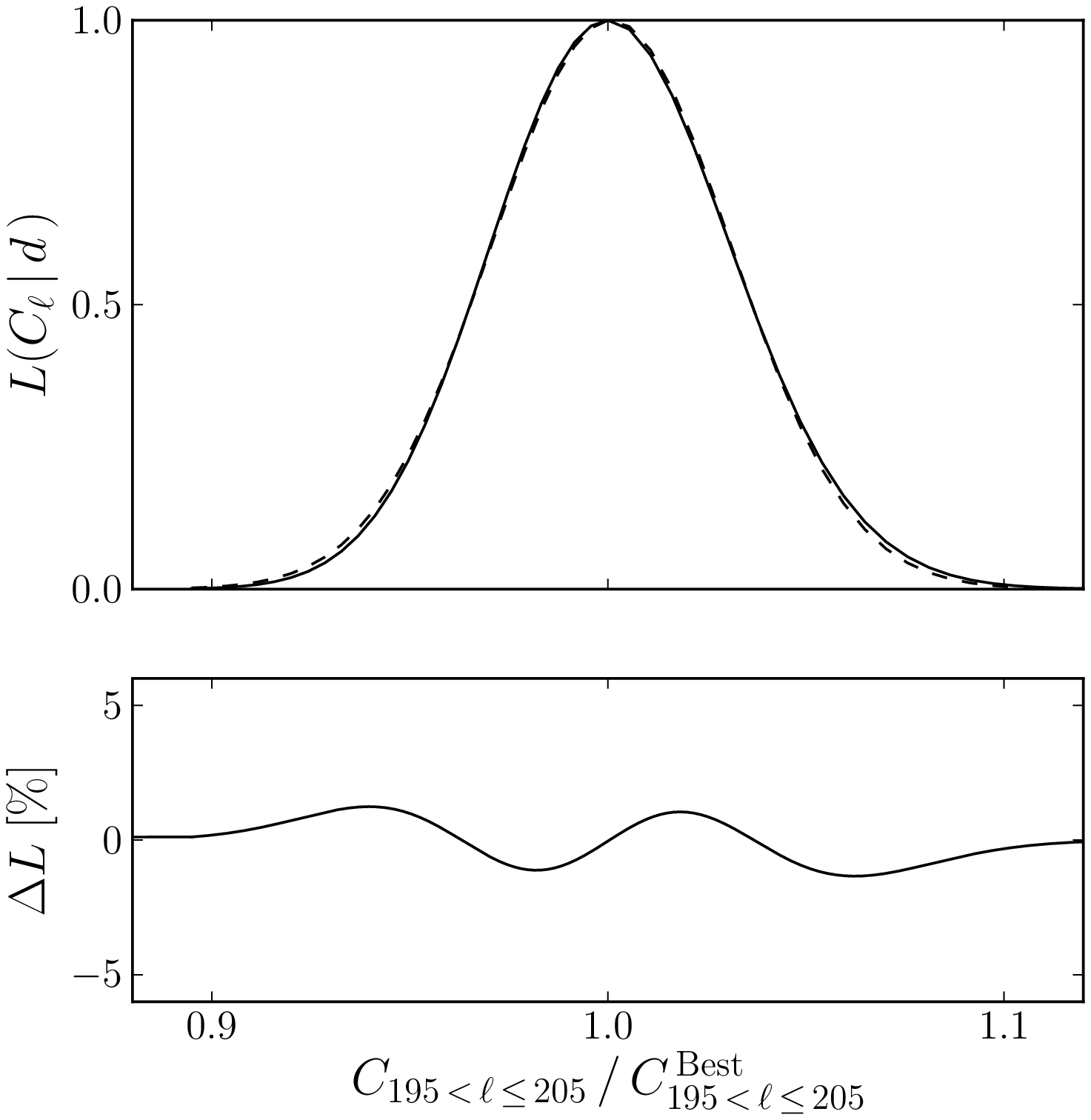}{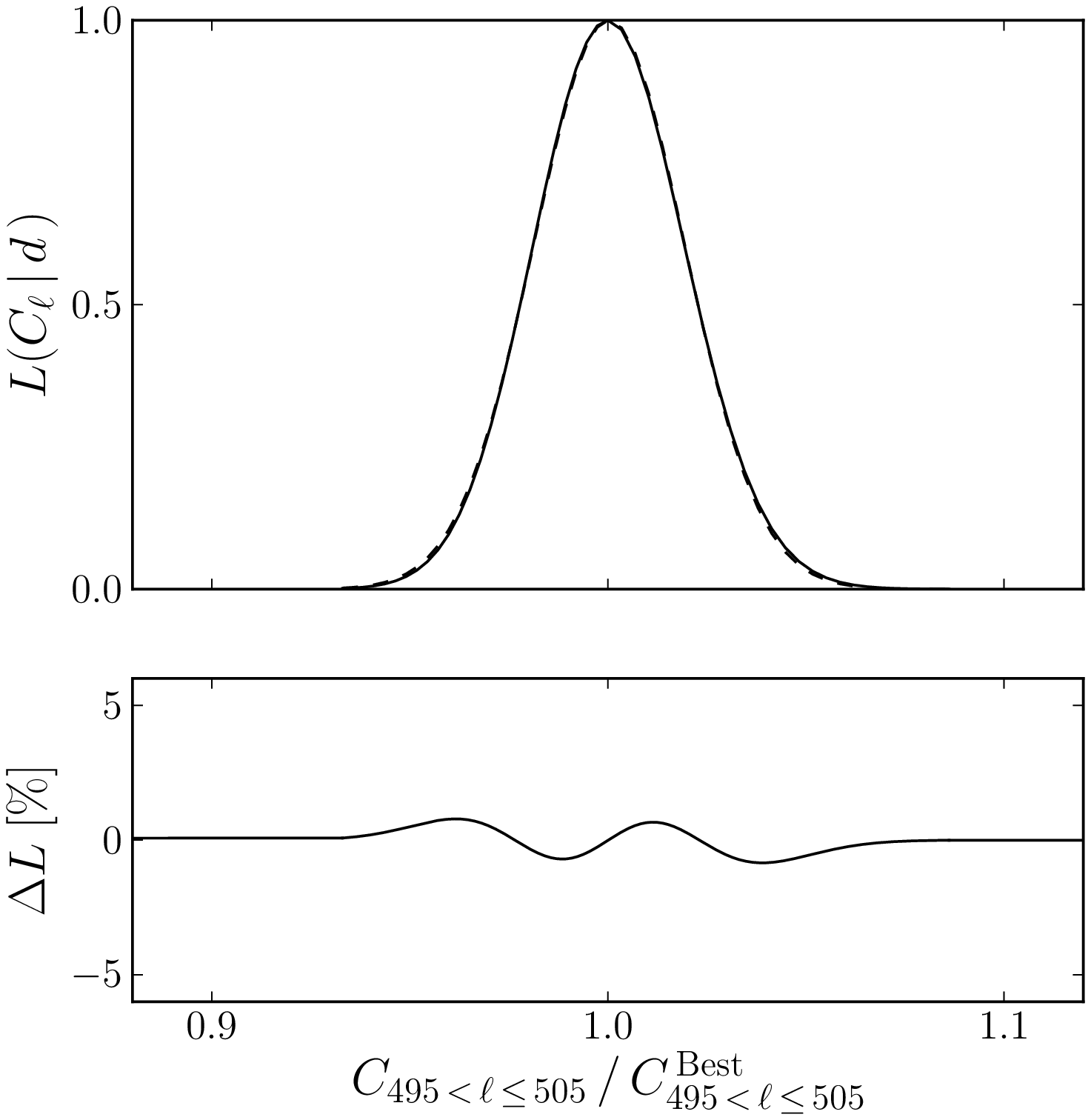}
{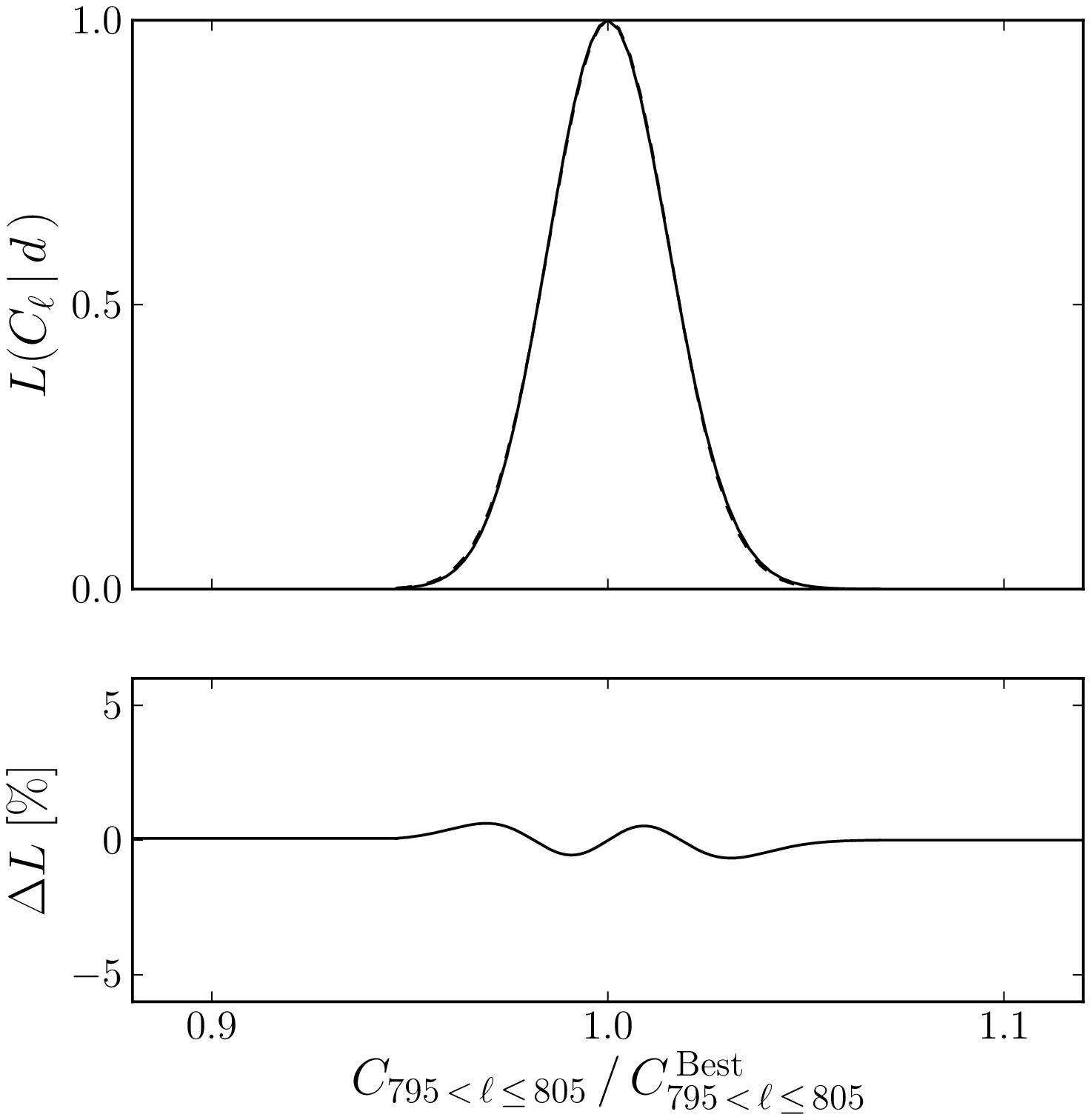}{fig:bin_like}
{Same as \fig{fig:bench_like}, but for a binned power
  spectrum. Combining adjacent power spectrum coefficients results in
  a more Gaussian probability distribution.}

\section{Discussion and conclusions}
\label{sec:summary}

Owing to its reduced size compared to a full data map, the power
spectrum of the CMB radiation anisotropies is a convenient
intermediate stage product in the process of estimating cosmological
parameters. Sufficient to completely characterize a Gaussian random
field, it contains all of the most fundamental cosmological
information present in the data \citep{1996PhRvD..54.1332J,
  1997MNRAS.291L..33B}. To construct the probability distributions of
the parameters out of the power spectrum requires a precise modeling
of their likelihood function. That is, one has to specify the shape of
the function, and the correlation between the individual power
spectrum coefficients.

Systematic effects, present in all real-world experiments, have the
potential to reduce the information content of a data set. In the
context of a CMB power spectrum analysis, issues may arise from beam
imperfections, incomplete sky coverage, and correlated noise. Here, we
conducted a systematic study of the impact of these effects on the
ability to constrain the CMB power spectrum. Assuming a simplified
scanning strategy that allows for an exact analytical treatment of the
problem, we used the Fisher matrix formalism for a quantitative
comparison of the information content of different experimental
setups. To complement the analysis, we also explored the shape of the
likelihood function directly by varying the amplitude of individual
power spectrum coefficients. We obtained the following results:

\begin{itemize}
\item For a high S/N experiment that covers 60~\% of the sky, we
  found neighboring power spectrum coefficients to be negatively
  correlated at the 10~\% level. Interestingly, the shape of the
  likelihood function deviated noticeably from a Gaussian
  distribution, typically by 2 - 5~\%. This was true even for the
  highest multipole values probed in our analysis, $\ell > 1000$.
\item Increasing the noise contribution changed the correlation
  structure between different power spectrum coefficients. It became
  somewhat more pronounced and of longer range. The residuals of a fit
  of the likelihood function with a Gaussian distribution remained
  basically the same.
\item Treated exactly, the mere presence of correlated noise did not
  lead to a loss of information. Comparing two simulations with white
  and colored noise power spectrum, we found $1/f^2$-type noise to be
  efficiently removed. In addition, it did not increase the amount of
  correlation among the power spectrum coefficients.
\item As expected, the impact of a reduced sky coverage on the
  analysis was most important. To study this effect, we further
  limited the observed region to 40~\% of the sky. As a result, the
  negative correlation between power spectrum coefficients increased
  to up to 18~\%. The strength of deviation of the likelihood function
  from Gaussianity, on the other hand, was mainly unaffected.
\item If power spectrum coefficients were combined into bins, the
  correlation structure among them simplified. Typically, only
  adjacent bins remained correlated at a noticeable level. Furthermore,
  the shape of the likelihood function became more Gaussian.
\item Finally, if asymmetric beams were present but could be accounted
  for exactly in the analysis, the Fisher matrix remained virtually
  unchanged compared to an equivalent experiment with symmetric
  beams. That is, asymmetric beams by themselves do not destroy
  information.
\end{itemize}

As a \emph{caveat lector}, we re-iterate that we calculate an upper
bound on the information content provided in these different
experimental scenarios assuming that all instrument properties are
known well enough that any small remaining uncertainty does not affect
these results. This reveals that the quality of a CMB power spectrum
measurement does not depend in principle on symmetric beams or the
absence of correlated noise. Recovering the information may require
sophisticated analysis and sufficient knowledge of the instrument
parameters. Nevertheless, this finding might inspire novel,
cost-effective experimental designs that embrace beam asymmetries
and/or correlated noise as long as these effects are well-calibrated.
It would be interesting in follow-up work to quantify the maximally
allowable calibration uncertainties for these statements to hold.

On a more practical level our results are useful in analysis of real
data sets in several ways: first we point out the level of
non-Gaussianity in the power spectrum likelihood that remains even for
relatively large sky coverage and at high $\ell$. We also illustrate
the typical correlation structure for measurements with an anisotropic
noise distribution. Third, these benchmarks help determine the point
of diminishing returns, i.e.\ when it no longer pays off to refine the
data analysis of a real data set. That point is reached when the
information recovered from the data set reaches a substantial fraction
of the theoretical information bound we give here.

\begin{acknowledgements}
  We are grateful to F. Bouchet, J-F. Cardoso, K. Benabed, and
  E. Hivon for useful discussions. BDW was supported by the ANR Chaire
  d'Excellence and NSF grants AST 07-08849 and AST 09-08902 during
  this work. FE gratefully acknowledges funding by the CNRS.
\end{acknowledgements}

\bibliographystyle{aa}
\bibliography{literature}

\begin{thebibliography}{42}
\expandafter\ifx\csname natexlab\endcsname\relax\def\natexlab#1{#1}\fi

\bibitem[{{Bennett} {et~al.}(2003){Bennett}, {Halpern}, {Hinshaw}, {Jarosik},
  {Kogut}, {Limon}, {Meyer}, {Page}, {Spergel}, {Tucker}, {Wollack}, {Wright},
  {Barnes}, {Greason}, {Hill}, {Komatsu}, {Nolta}, {Odegard}, {Peiris},
  {Verde}, \& {Weiland}}]{2003ApJS..148....1B}
{Bennett}, C.~L., {Halpern}, M., {Hinshaw}, G., {et~al.} 2003, \apjs, 148, 1

\bibitem[{{Berti} {et~al.}(2005){Berti}, {Buonanno}, \&
  {Will}}]{2005PhRvD..71h4025B}
{Berti}, E., {Buonanno}, A., \& {Will}, C.~M. 2005, \prd, 71, 084025

\bibitem[{{Bond} {et~al.}(1997){Bond}, {Efstathiou}, \&
  {Tegmark}}]{1997MNRAS.291L..33B}
{Bond}, J.~R., {Efstathiou}, G., \& {Tegmark}, M. 1997, \mnras, 291, L33

\bibitem[{{Bond} {et~al.}(1998){Bond}, {Jaffe}, \&
  {Knox}}]{1998PhRvD..57.2117B}
{Bond}, J.~R., {Jaffe}, A.~H., \& {Knox}, L. 1998, \prd, 57, 2117

\bibitem[{{Borrill}(1999)}]{1999PhRvD..59b7302B}
{Borrill}, J. 1999, \prd, 59, 027302

\bibitem[{{Burigana} \& {S{\'a}ez}(2003)}]{2003A&A...409..423B}
{Burigana}, C. \& {S{\'a}ez}, D. 2003, \aap, 409, 423

\bibitem[{{Christensen} \& {Meyer}(2000)}]{2000astro.ph..6401C}
{Christensen}, N. \& {Meyer}, R. 2000, ArXiv Astrophysics e-prints

\bibitem[{{Delabrouille}(1998)}]{1998A&AS..127..555D}
{Delabrouille}, J. 1998, \aaps, 127, 555

\bibitem[{{Dunkley} {et~al.}(2011){Dunkley}, {Hlozek}, {Sievers}, {Acquaviva},
  {Ade}, {Aguirre}, {Amiri}, {Appel}, {Barrientos}, {Battistelli}, {Bond},
  {Brown}, {Burger}, {Chervenak}, {Das}, {Devlin}, {Dicker}, {Bertrand
  Doriese}, {D{\"u}nner}, {Essinger-Hileman}, {Fisher}, {Fowler}, {Hajian},
  {Halpern}, {Hasselfield}, {Hern{\'a}ndez-Monteagudo}, {Hilton}, {Hilton},
  {Hincks}, {Huffenberger}, {Hughes}, {Hughes}, {Infante}, {Irwin}, {Juin},
  {Kaul}, {Klein}, {Kosowsky}, {Lau}, {Limon}, {Lin}, {Lupton}, {Marriage},
  {Marsden}, {Mauskopf}, {Menanteau}, {Moodley}, {Moseley}, {Netterfield},
  {Niemack}, {Nolta}, {Page}, {Parker}, {Partridge}, {Reid}, {Sehgal},
  {Sherwin}, {Spergel}, {Staggs}, {Swetz}, {Switzer}, {Thornton}, {Trac},
  {Tucker}, {Warne}, {Wollack}, \& {Zhao}}]{2011ApJ...739...52D}
{Dunkley}, J., {Hlozek}, R., {Sievers}, J., {et~al.} 2011, \apj, 739, 52

\bibitem[{{Efstathiou}(2004)}]{2004MNRAS.349..603E}
{Efstathiou}, G. 2004, \mnras, 349, 603

\bibitem[{{Efstathiou} {et~al.}(2005){Efstathiou}, {Lawrence}, \&
  J.}]{2005planck_bluebook}
{Efstathiou}, G., {Lawrence}, C., \& J., T. 2005, {Planck. The Scientific
  Program}, Planck Science Team

\bibitem[{{Elsner} \& {Wandelt}(2012)}]{2012A&A...540L...6E}
{Elsner}, F. \& {Wandelt}, B.~D. 2012, \aap, 540, L6

\bibitem[{{Gruppuso} {et~al.}(2009){Gruppuso}, {de Rosa}, {Cabella}, {Paci},
  {Finelli}, {Natoli}, {de Gasperis}, \& {Mandolesi}}]{2009MNRAS.400..463G}
{Gruppuso}, A., {de Rosa}, A., {Cabella}, P., {et~al.} 2009, \mnras, 400, 463

\bibitem[{{Hinshaw} {et~al.}(2007){Hinshaw}, {Nolta}, {Bennett}, {Bean},
  {Dor{\'e}}, {Greason}, {Halpern}, {Hill}, {Jarosik}, {Kogut}, {Komatsu},
  {Limon}, {Odegard}, {Meyer}, {Page}, {Peiris}, {Spergel}, {Tucker}, {Verde},
  {Weiland}, {Wollack}, \& {Wright}}]{2007ApJS..170..288H}
{Hinshaw}, G., {Nolta}, M.~R., {Bennett}, C.~L., {et~al.} 2007, \apjs, 170, 288

\bibitem[{{Hinshaw} {et~al.}(2009){Hinshaw}, {Weiland}, {Hill}, {Odegard},
  {Larson}, {Bennett}, {Dunkley}, {Gold}, {Greason}, {Jarosik}, {Komatsu},
  {Nolta}, {Page}, {Spergel}, {Wollack}, {Halpern}, {Kogut}, {Limon}, {Meyer},
  {Tucker}, \& {Wright}}]{2009ApJS..180..225H}
{Hinshaw}, G., {Weiland}, J.~L., {Hill}, R.~S., {et~al.} 2009, \apjs, 180, 225

\bibitem[{{Hivon} {et~al.}(2002){Hivon}, {G{\'o}rski}, {Netterfield}, {Crill},
  {Prunet}, \& {Hansen}}]{2002ApJ...567....2H}
{Hivon}, E., {G{\'o}rski}, K.~M., {Netterfield}, C.~B., {et~al.} 2002, \apj,
  567, 2

\bibitem[{{Hu} \& {Tegmark}(1999)}]{1999ApJ...514L..65H}
{Hu}, W. \& {Tegmark}, M. 1999, \apjl, 514, L65

\bibitem[{{Jarosik} {et~al.}(2011){Jarosik}, {Bennett}, {Dunkley}, {Gold},
  {Greason}, {Halpern}, {Hill}, {Hinshaw}, {Kogut}, {Komatsu}, {Larson},
  {Limon}, {Meyer}, {Nolta}, {Odegard}, {Page}, {Smith}, {Spergel}, {Tucker},
  {Weiland}, {Wollack}, \& {Wright}}]{2011ApJS..192...14J}
{Jarosik}, N., {Bennett}, C.~L., {Dunkley}, J., {et~al.} 2011, \apjs, 192, 14

\bibitem[{{Jungman} {et~al.}(1996){Jungman}, {Kamionkowski}, {Kosowsky}, \&
  {Spergel}}]{1996PhRvD..54.1332J}
{Jungman}, G., {Kamionkowski}, M., {Kosowsky}, A., \& {Spergel}, D.~N. 1996,
  \prd, 54, 1332

\bibitem[{{Keisler} {et~al.}(2011){Keisler}, {Reichardt}, {Aird}, {Benson},
  {Bleem}, {Carlstrom}, {Chang}, {Cho}, {Crawford}, {Crites}, {de Haan},
  {Dobbs}, {Dudley}, {George}, {Halverson}, {Holder}, {Holzapfel}, {Hoover},
  {Hou}, {Hrubes}, {Joy}, {Knox}, {Lee}, {Leitch}, {Lueker}, {Luong-Van},
  {McMahon}, {Mehl}, {Meyer}, {Millea}, {Mohr}, {Montroy}, {Natoli}, {Padin},
  {Plagge}, {Pryke}, {Ruhl}, {Schaffer}, {Shaw}, {Shirokoff}, {Spieler},
  {Staniszewski}, {Stark}, {Story}, {van Engelen}, {Vanderlinde}, {Vieira},
  {Williamson}, \& {Zahn}}]{2011ApJ...743...28K}
{Keisler}, R., {Reichardt}, C.~L., {Aird}, K.~A., {et~al.} 2011, \apj, 743, 28

\bibitem[{{Komatsu} {et~al.}(2011){Komatsu}, {Smith}, {Dunkley}, {Bennett},
  {Gold}, {Hinshaw}, {Jarosik}, {Larson}, {Nolta}, {Page}, {Spergel},
  {Halpern}, {Hill}, {Kogut}, {Limon}, {Meyer}, {Odegard}, {Tucker}, {Weiland},
  {Wollack}, \& {Wright}}]{2011ApJS..192...18K}
{Komatsu}, E., {Smith}, K.~M., {Dunkley}, J., {et~al.} 2011, \apjs, 192, 18

\bibitem[{{Kosowsky}(2003)}]{2003NewAR..47..939K}
{Kosowsky}, A. 2003, \nar, 47, 939

\bibitem[{{Kuo} {et~al.}(2004){Kuo}, {Ade}, {Bock}, {Cantalupo}, {Daub},
  {Goldstein}, {Holzapfel}, {Lange}, {Lueker}, {Newcomb}, {Peterson}, {Ruhl},
  {Runyan}, \& {Torbet}}]{2004ApJ...600...32K}
{Kuo}, C.~L., {Ade}, P.~A.~R., {Bock}, J.~J., {et~al.} 2004, \apj, 600, 32

\bibitem[{{Lee} \& {Pen}(2008)}]{2008ApJ...686L...1L}
{Lee}, J. \& {Pen}, U.-L. 2008, \apjl, 686, L1

\bibitem[{{Lewis} \& {Bridle}(2002)}]{2002PhRvD..66j3511L}
{Lewis}, A. \& {Bridle}, S. 2002, \prd, 66, 103511

\bibitem[{{Maino} {et~al.}(1999){Maino}, {Burigana}, {Maltoni}, {Wandelt},
  {G{\'o}rski}, {Malaspina}, {Bersanelli}, {Mandolesi}, {Banday}, \&
  {Hivon}}]{1999A&AS..140..383M}
{Maino}, D., {Burigana}, C., {Maltoni}, M., {et~al.} 1999, \aaps, 140, 383

\bibitem[{{Mennella} {et~al.}(2011){Mennella}, {Butler}, {Curto}, {Cuttaia},
  {Davis}, {Dick}, {Frailis}, {Galeotta}, {Gregorio}, {Kurki-Suonio},
  {Lawrence}, {Leach}, {Leahy}, {Lowe}, {Maino}, {Mandolesi}, {Maris},
  {Mart{\'{\i}}nez-Gonz{\'a}lez}, {Meinhold}, {Morgante}, {Pearson},
  {Perrotta}, {Polenta}, {Poutanen}, {Sandri}, {Seiffert}, {Suur-Uski},
  {Tavagnacco}, {Terenzi}, {Tomasi}, {Valiviita}, {Villa}, {Watson},
  {Wilkinson}, {Zacchei}, {Zonca}, {Aja}, {Artal}, {Baccigalupi}, {Banday},
  {Barreiro}, {Bartlett}, {Bartolo}, {Battaglia}, {Bennett}, {Bonaldi},
  {Bonavera}, {Borrill}, {Bouchet}, {Burigana}, {Cabella}, {Cappellini},
  {Chen}, {Colombo}, {Cruz}, {Danese}, {D'Arcangelo}, {Davies}, {de Gasperis},
  {de Rosa}, {de Zotti}, {Dickinson}, {Diego}, {Donzelli}, {Efstathiou},
  {En{\ss}lin}, {Eriksen}, {Falvella}, {Finelli}, {Foley}, {Franceschet},
  {Franceschi}, {Gaier}, {G{\'e}nova-Santos}, {George}, {G{\'o}mez},
  {Gonz{\'a}lez-Nuevo}, {G{\'o}rski}, {Gruppuso}, {Hansen}, {Herranz},
  {Herreros}, {Hoyland}, {Hughes}, {Jewell}, {Jukkala}, {Juvela},
  {Kangaslahti}, {Keih{\"a}nen}, {Keskitalo}, {Kilpia}, {Kisner}, {Knoche},
  {Knox}, {Laaninen}, {L{\"a}hteenm{\"a}ki}, {Lamarre}, {Leonardi},
  {Le{\'o}n-Tavares}, {Leutenegger}, {Lilje}, {L{\'o}pez-Caniego}, {Lubin},
  {Malaspina}, {Marinucci}, {Massardi}, {Matarrese}, {Matthai}, {Melchiorri},
  {Mendes}, {Miccolis}, {Migliaccio}, {Mitra}, {Moss}, {Natoli}, {Nesti},
  {N{\o}rgaard-Nielsen}, {Pagano}, {Paladini}, {Paoletti}, {Partridge},
  {Pasian}, {Pettorino}, {Pietrobon}, {Pospieszalski}, {Pr{\'e}zeau}, {Prina},
  {Procopio}, {Puget}, {Quercellini}, {Rachen}, {Rebolo}, {Reinecke},
  {Ricciardi}, {Robbers}, {Rocha}, {Roddis}, {Rubino-Mart{\'{\i}}n},
  {Savelainen}, {Scott}, {Silvestri}, {Simonetto}, {Sjoman}, {Smoot}, {Sozzi},
  {Stringhetti}, {Tauber}, {Tofani}, {Toffolatti}, {Tuovinen}, {T{\"u}rler},
  {Umana}, {Valenziano}, {Varis}, {Vielva}, {Vittorio}, {Wade}, {Watson},
  {White}, \& {Winder}}]{2011A&A...536A...3M}
{Mennella}, A., {Butler}, R.~C., {Curto}, A., {et~al.} 2011, \aap, 536, A3

\bibitem[{{Mitra} {et~al.}(2011){Mitra}, {Rocha}, {G{\'o}rski}, {Huffenberger},
  {Eriksen}, {Ashdown}, \& {Lawrence}}]{2011ApJS..193....5M}
{Mitra}, S., {Rocha}, G., {G{\'o}rski}, K.~M., {et~al.} 2011, \apjs, 193, 5

\bibitem[{{Mukhanov}(2004)}]{2004IJTP...43..623M}
{Mukhanov}, V. 2004, International Journal of Theoretical Physics, 43, 623

\bibitem[{{Netterfield} {et~al.}(2002){Netterfield}, {Ade}, {Bock}, {Bond},
  {Borrill}, {Boscaleri}, {Coble}, {Contaldi}, {Crill}, {de Bernardis},
  {Farese}, {Ganga}, {Giacometti}, {Hivon}, {Hristov}, {Iacoangeli}, {Jaffe},
  {Jones}, {Lange}, {Martinis}, {Masi}, {Mason}, {Mauskopf}, {Melchiorri},
  {Montroy}, {Pascale}, {Piacentini}, {Pogosyan}, {Pongetti}, {Prunet},
  {Romeo}, {Ruhl}, \& {Scaramuzzi}}]{2002ApJ...571..604N}
{Netterfield}, C.~B., {Ade}, P.~A.~R., {Bock}, J.~J., {et~al.} 2002, \apj, 571,
  604

\bibitem[{{Oh} {et~al.}(1999){Oh}, {Spergel}, \&
  {Hinshaw}}]{1999ApJ...510..551O}
{Oh}, S.~P., {Spergel}, D.~N., \& {Hinshaw}, G. 1999, \apj, 510, 551

\bibitem[{{Planck Collaboration} {et~al.}(2011){Planck Collaboration}, {Ade},
  {Aghanim}, {Arnaud}, {Ashdown}, {Aumont}, {Baccigalupi}, {Baker}, {Balbi},
  {Banday}, \& et~al.}]{2011A&A...536A...1P}
{Planck Collaboration}, {Ade}, P.~A.~R., {Aghanim}, N., {et~al.} 2011, \aap,
  536, A1

\bibitem[{{Planck HFI Core Team} {et~al.}(2011){Planck HFI Core Team}, {Ade},
  {Aghanim}, {Ansari}, {Arnaud}, {Ashdown}, {Aumont}, {Banday}, {Bartelmann},
  {Bartlett}, {Battaner}, {Benabed}, {Beno{\^i}t}, {Bernard}, {Bersanelli},
  {Bhatia}, {Bock}, {Bond}, {Borrill}, {Bouchet}, {Boulanger}, {Bradshaw},
  {Br{\'e}elle}, {Bucher}, {Camus}, {Cardoso}, {Catalano}, {Challinor},
  {Chamballu}, {Charra}, {Charra}, {Chary}, {Chiang}, {Church}, {Clements},
  {Colombi}, {Couchot}, {Coulais}, {Cressiot}, {Crill}, {Crook}, {de
  Bernardis}, {Delabrouille}, {Delouis}, {D{\'e}sert}, {Dolag}, {Dole},
  {Dor{\'e}}, {Douspis}, {Efstathiou}, {Eng}, {Filliard}, {Forni}, {Fosalba},
  {Fourmond}, {Ganga}, {Giard}, {Girard}, {Giraud-H{\'e}raud}, {Gispert},
  {G{\'o}rski}, {Gratton}, {Griffin}, {Guyot}, {Haissinski}, {Harrison},
  {Helou}, {Henrot-Versill{\'e}}, {Hern{\'a}ndez-Monteagudo}, {Hildebrandt},
  {Hills}, {Hivon}, {Hobson}, {Holmes}, {Huffenberger}, {Jaffe}, {Jones},
  {Kaplan}, {Kneissl}, {Knox}, {Lagache}, {Lamarre}, {Lami}, {Lange},
  {Lasenby}, {Lavabre}, {Lawrence}, {Leriche}, {Leroy}, {Longval},
  {Mac{\'{\i}}as-P{\'e}rez}, {Maciaszek}, {MacTavish}, {Maffei}, {Mandolesi},
  {Mann}, {Mansoux}, {Masi}, {Matsumura}, {McGehee}, {Melin}, {Mercier},
  {Miville-Desch{\^e}nes}, {Moneti}, {Montier}, {Mortlock}, {Murphy}, {Nati},
  {Netterfield}, {N{\o}rgaard-Nielsen}, {North}, {Noviello}, {Novikov},
  {Osborne}, {Paine}, {Pajot}, {Patanchon}, {Peacocke}, {Pearson}, {Perdereau},
  {Perotto}, {Piacentini}, {Piat}, {Plaszczynski}, {Pointecouteau}, {Pons},
  {Ponthieu}, {Pr{\'e}zeau}, {Prunet}, {Puget}, {Reach}, {Renault},
  {Ristorcelli}, {Rocha}, {Rosset}, {Roudier}, {Rowan-Robinson}, {Rusholme},
  {Santos}, {Savini}, {Schaefer}, {Shellard}, {Spencer}, {Starck}, {Stassi},
  {Stolyarov}, {Stompor}, {Sudiwala}, {Sunyaev}, {Sygnet}, {Tauber}, {Thum},
  {Torre}, {Touze}, {Tristram}, {van Leeuwen}, {Vibert}, {Vibert}, {Wade},
  {Wandelt}, {White}, {Wiesemeyer}, {Woodcraft}, {Yurchenko}, {Yvon}, \&
  {Zacchei}}]{2011A&A...536A...4P}
{Planck HFI Core Team}, {Ade}, P.~A.~R., {Aghanim}, N., {et~al.} 2011, \aap,
  536, A4

\bibitem[{{Rimes} \& {Hamilton}(2005)}]{2005MNRAS.360L..82R}
{Rimes}, C.~D. \& {Hamilton}, A.~J.~S. 2005, \mnras, 360, L82

\bibitem[{{Ruhl} {et~al.}(2004){Ruhl}, {Ade}, {Carlstrom}, {Cho}, {Crawford},
  {Dobbs}, {Greer}, {Halverson}, {Holzapfel}, {Lanting}, {Lee}, {Leitch},
  {Leong}, {Lu}, {Lueker}, {Mehl}, {Meyer}, {Mohr}, {Padin}, {Plagge}, {Pryke},
  {Runyan}, {Schwan}, {Sharp}, {Spieler}, {Staniszewski}, \&
  {Stark}}]{2004SPIE.5498...11R}
{Ruhl}, J., {Ade}, P.~A.~R., {Carlstrom}, J.~E., {et~al.} 2004, in Presented at
  the Society of Photo-Optical Instrumentation Engineers (SPIE) Conference,
  Vol. 5498, Society of Photo-Optical Instrumentation Engineers (SPIE)
  Conference Series, ed. {C.~M.~Bradford, P.~A.~R.~Ade, J.~E.~Aguirre,
  J.~J.~Bock, M.~Dragovan, L.~Duband, L.~Earle, J.~Glenn, H.~Matsuhara,
  B.~J.~Naylor, H.~T.~Nguyen, M.~Yun, \& J.~Zmuidzinas}, 11--29

\bibitem[{{Smoot} {et~al.}(1992){Smoot}, {Bennett}, {Kogut}, {Wright}, {Aymon},
  {Boggess}, {Cheng}, {de Amici}, {Gulkis}, {Hauser}, {Hinshaw}, {Jackson},
  {Janssen}, {Kaita}, {Kelsall}, {Keegstra}, {Lineweaver}, {Loewenstein},
  {Lubin}, {Mather}, {Meyer}, {Moseley}, {Murdock}, {Rokke}, {Silverberg},
  {Tenorio}, {Weiss}, \& {Wilkinson}}]{1992ApJ...396L...1S}
{Smoot}, G.~F., {Bennett}, C.~L., {Kogut}, A., {et~al.} 1992, \apjl, 396, L1

\bibitem[{{Szapudi} {et~al.}(2001){Szapudi}, {Prunet}, \&
  {Colombi}}]{2001ApJ...561L..11S}
{Szapudi}, I., {Prunet}, S., \& {Colombi}, S. 2001, \apjl, 561, L11

\bibitem[{{Tegmark}(1997)}]{1997PhRvD..55.5895T}
{Tegmark}, M. 1997, \prd, 55, 5895

\bibitem[{{Verde} {et~al.}(2003){Verde}, {Peiris}, {Spergel}, {Nolta},
  {Bennett}, {Halpern}, {Hinshaw}, {Jarosik}, {Kogut}, {Limon}, {Meyer},
  {Page}, {Tucker}, {Wollack}, \& {Wright}}]{2003ApJS..148..195V}
{Verde}, L., {Peiris}, H.~V., {Spergel}, D.~N., {et~al.} 2003, \apjs, 148, 195

\bibitem[{{Wandelt} \& {Hansen}(2003)}]{2003PhRvD..67b3001W}
{Wandelt}, B.~D. \& {Hansen}, F.~K. 2003, \prd, 67, 023001

\bibitem[{{Wandelt} {et~al.}(2001){Wandelt}, {Hivon}, \&
  {G{\'o}rski}}]{2001PhRvD..64h3003W}
{Wandelt}, B.~D., {Hivon}, E., \& {G{\'o}rski}, K.~M. 2001, \prd, 64, 083003

\bibitem[{{Wandelt} {et~al.}(2004){Wandelt}, {Larson}, \&
  {Lakshminarayanan}}]{2004PhRvD..70h3511W}
{Wandelt}, B.~D., {Larson}, D.~L., \& {Lakshminarayanan}, A. 2004, \prd, 70,
  083511

\end{thebibliography}

\end{document}